\documentclass[
aip, sd,
amsmath, amssymb, superscriptaddress,
 reprint,
]{revtex4-2}

\usepackage{graphicx}% Include figure files
\usepackage{dcolumn}% Align table columns on decimal point
\usepackage{bm}% bold math

\usepackage[utf8]{inputenc}
\usepackage[T1]{fontenc}
\usepackage{mathptmx}

\usepackage[separate-uncertainty=true, range-units=repeat, range-phrase={\ to }]{siunitx}
\DeclareSIUnit\fluence{\milli\joule\per\centi\meter\squared}
\usepackage[version=4]{mhchem}
\usepackage{graphicx, float}
\usepackage[capitalise]{cleveref}

\begin{document}
\title{Pump--probe x-ray microscopy of photo-induced magnetization dynamics at MHz repetition rates}

\author{Kathinka Gerlinger} %experiment, data anlysis, manuscript
 \affiliation{Max Born Institute for Nonlinear Optics and Short Pulse Spectroscopy, 12489 Berlin, Germany}
\author{Bastian Pfau} %experimental design, exp. supervision, manuscript
 \email{bastian.pfau@mbi-berlin.de}
 \affiliation{Max Born Institute for Nonlinear Optics and Short Pulse Spectroscopy, 12489 Berlin, Germany}
\author{Martin Hennecke} %interpretation of the data analysis, manuscript
 \affiliation{Max Born Institute for Nonlinear Optics and Short Pulse Spectroscopy, 12489 Berlin, Germany}
\author{Lisa-Marie Kern} %experiment
 \affiliation{Max Born Institute for Nonlinear Optics and Short Pulse Spectroscopy, 12489 Berlin, Germany}
\author{Ingo Will} %experiment, laser
 \affiliation{Max Born Institute for Nonlinear Optics and Short Pulse Spectroscopy, 12489 Berlin, Germany}
\author{Tino Noll} %laser
 \affiliation{Max Born Institute for Nonlinear Optics and Short Pulse Spectroscopy, 12489 Berlin, Germany}
\author{Markus Weigand} %beamline scientist
 \affiliation{Helmholtz-Zentrum Berlin für Materialien und Energie, 12489 Berlin, Germany}
\author{Joachim Gräfe} %beamline scientist
 \affiliation{Max Planck Institute for Intelligent Systems, 70569 Stuttgart, Germany}
\author{Nick Träger} %experiment
 \affiliation{Max Planck Institute for Intelligent Systems, 70569 Stuttgart, Germany}
\author{Michael Schneider} %samples
 \affiliation{Max Born Institute for Nonlinear Optics and Short Pulse Spectroscopy, 12489 Berlin, Germany}
\author{Christian M. Günther} %samples?
 \affiliation{Technische Universität Berlin, Zentraleinrichtung Elektronenmikroskopie (ZELMI), 10623 Berlin, Germany}
\author{Dieter Engel} %samples
 \affiliation{Max Born Institute for Nonlinear Optics and Short Pulse Spectroscopy, 12489 Berlin, Germany}
\author{Gisela Schütz} %supervision MPI
 \affiliation{Max Planck Institute for Intelligent Systems, 70569 Stuttgart, Germany}
\author{Stefan Eisebitt} %supervision MBI
 \affiliation{Max Born Institute for Nonlinear Optics and Short Pulse Spectroscopy, 12489 Berlin, Germany}
 \affiliation{Technische Universität Berlin, Institut für Optik und Atomare Physik, 10623 Berlin, Germany}

\date{\today}

\begin{abstract}
We present time-resolved scanning x-ray microscopy measurements with picosecond photo-excitation via a tailored infrared pump laser at a scanning transmission x-ray microscope.
Specifically, we image the laser-induced demagnetization and remagnetization of thin ferrimagnetic GdFe films proceeding on a few nanoseconds time scale.
Controlling the heat load on the sample via additional reflector and heatsink layers allows us to conduct destruction-free measurements at a repetition rate of \SI{50}{\mega\hertz}.
Near-field enhancement of the photo-excitation and controlled annealing effects lead to laterally heterogeneous magnetization dynamics which we trace with \SI{30}{\nano\meter} spatial resolution.
Our work opens new opportunities to study photo-induced dynamics on the nanometer scale, with access to picosecond to nanosecond time scales, which is of technological relevance, especially in the field of magnetism. 

\end{abstract}
\maketitle

\section{Introduction}
Studying magnetization dynamics on the nanometer scale is important for our fundamental understanding of magnetic materials and their use in technological applications.\cite{Vedmedenko2020,kent2020} 
Soft-x-ray microscopes using synchrotron radiation (SR) can resolve these dynamics down to a few nanometers with element specificity employing the x-ray magnetic circular dichroism (XMCD) as magnetic contrast mechanism.\cite{stoll2015,fischer2017}
The time-resolution of these microscopes is limited, on the one hand, by the duration of the SR bunches to typically \SIrange{30}{100}{ps}, depending on the storage ring and the mode of operation.
On the other hand, additional restrictions arise from the pulse duration of the excitation, which, today, is almost exclusively realized electrically, typically on a time scale longer than \SI{100}{\pico\second}.
Ultrashort laser pulses are able to trigger magnetization dynamics down to the pico- or even femtosecond regime and provide access to photo-induced magnetization phenomena, such as, e.g., ultrafast demagnetization and all-optical magnetization switching,\cite{Kirilyuk2010,Carva2017} topological phase transitions,\cite{Buttner2021, Gerlinger2021} optical spin wave excitation\cite{Subkhangulov2015,Satoh2012} or the control of spin torques.\cite{Schellekens2014,Kim2015}
In addition, laser pulses can also be used to generate significantly shorter electrical excitation pulses than typically feasible with purely electronic signal generators.\cite{Choe2004,Heyne2010}
Therefore, the integration of short-pulsed lasers into x-ray microscopes at SR sources pushes these nanometer-scale imaging instruments to better time-resolution and extends their scope towards sub-nanosecond magnetization dynamics that are only accessible via ultrashort photo-excitation. 

While lasers with high repetition rates have already been employed for studying magnetization dynamics with photo-emission electron microscopy (PEEM),\cite{LeGuyader2015,Gierster2015} they have not yet been implemented in a scanning transmission x-ray microscope (STXM).
One reason is the typically extremely limited space between the x-ray lenses and the sample, which makes the integration of a pump laser beam into a STXM very challenging.
Furthermore, the laser must be synchronized to the bunch clock of the storage ring, which is particularly demanding when, in addition, gapless tunability of the pump repetition rate is needed for the highly efficient asynchronous excitation schemes used for time-resolved measurements in modern STXMs.\cite{Bonetti2015, Weigand2022}

Complying with the orbital frequency of the storage ring and ensuring reasonable data collection times, time-resolved imaging at SR sources is best performed at \si{\mega\hertz} repetition rates. 
However, at such high repetition rates, the extreme thermal load on solid-state samples poses an additional dilemma to laser-based pump--probe experiments as it can change the sample's magnetic properties and, ultimately, lead to permanent degradation or even destruction.
In general, the heat load on the sample can be reduced by shrinking the spot size of the pump beam, maintaining an identical excitation fluence with lower average power incident on the sample.
However, in order to probe an almost homogeneously pumped area, the probe beam size has to be reduced accordingly.
In this work, we use a STXM to probe our samples, which is based on an x-ray beam ultimately focused down to a few nanometers.
This technique, therefore, allows reducing the effectively pumped area to micrometer size and, in this way, minimizing the heat load.
As a result, we demonstrate destruction-free time-resolved imaging at pump--probe repetition rates of up to \SI{50}{\mega\hertz} at fluences exceeding \SI{3}{\fluence}.
The photo-excitation is realized using a custom-developed infrared (IR) pump laser newly installed at the MAXYMUS microscope---a STXM at the SR facility BESSY II (Berlin, Germany).

We apply this pump--probe imaging method to detect ultrafast photo-induced demagnetization and magnetization recovery in GdFe alloys with \SI{30}{\nano\meter} spatial and down to \SI{50}{\pico\second} temporal resolution in standard multi-bunch operation mode at BESSY II. 
Rare-earth transition metal alloys such as GdFe are known for their intriguing all-optical magnetization switching phenomena,\cite{Stanciu2007, Radu2011} making these materials attractive for potential data storage applications, as well as a model system for distinct multisublattice spin and orbital moment dynamics.\cite{Bergeard2014, Radu2015, Hennecke2019}
As detailed below, time-resolved imaging with high spatial resolution allows us to reveal nanometer-sized areas created by the pump laser that show very different magnetization dynamics compared to the rest of the film.
Our results demonstrate that STXM is a highly valuable tool to image photo-induced pico- to nanosecond magnetization dynamics on the nanometer scale, which is of high relevance in view of future spintronics applications, e.g., to study the dynamics of photo-induced magnetic texture formation\cite{Buttner2021,Steinbach2022} or collective excitations.\cite{Satoh2012}

\section{Experimental Details}

The experiments were conducted using Ta(3)/Gd$_{29}$Fe$_{71}$(20)/Pt(3) (thicknesses in \si{\nano\meter}) and Ta(3)/Gd$_{27}$Fe$_{73}$(20)/Pt(3) films, deposited on \SI{150}{nm} thin silicon-nitride membranes via magnetron sputtering.  
The films exhibit a perpendicular magnetic anisotropy and a typical square-shaped out-of-plane magnetic hysteresis with a coercive field of \SI{30}{mT} and \SI{45}{mT}, respectively.
Magnetic-contrast images of the samples were acquired with circularly polarized soft x-rays tuned to the Gd M$_5$ edge (\SI{1190}{\eV}), providing sensitivity to the out-of-plane magnetization component of the Gd sublattice. 
We record a complete XMCD data set ($I_+(x,y)$, $I_-(x,y)$) by switching either the helicity of the x-ray probe or the polarity of the applied out-of-plane field of \SI{200}{\milli\tesla}, provided by four rotatable permanent magnets.\cite{Nolle2012} 
The normalized out-of-plane magnetization $m_z(x,y)$ is retrieved from the XMCD contrast $D(x,y)$ which in turn is calculated from the two images ($I_+$, $I_-$) by $D(x,y) = \log \left( I_+(x,y) / I_-(x,y) \right)$. Each XMCD image is then normalized to a pixel row at the edge of the scanning area, which is unaffected by the laser excitation. This procedure also corrects small intensity line artifacts from the scanning mode of operation.
The scanning field of view is \SIrange[range-phrase=$\times$, range-units=single]{2.4}{2.4}{\micro\meter\squared} with a resolution of \SI{30}{\nano\meter}.

\begin{figure}
    \centering
    \includegraphics{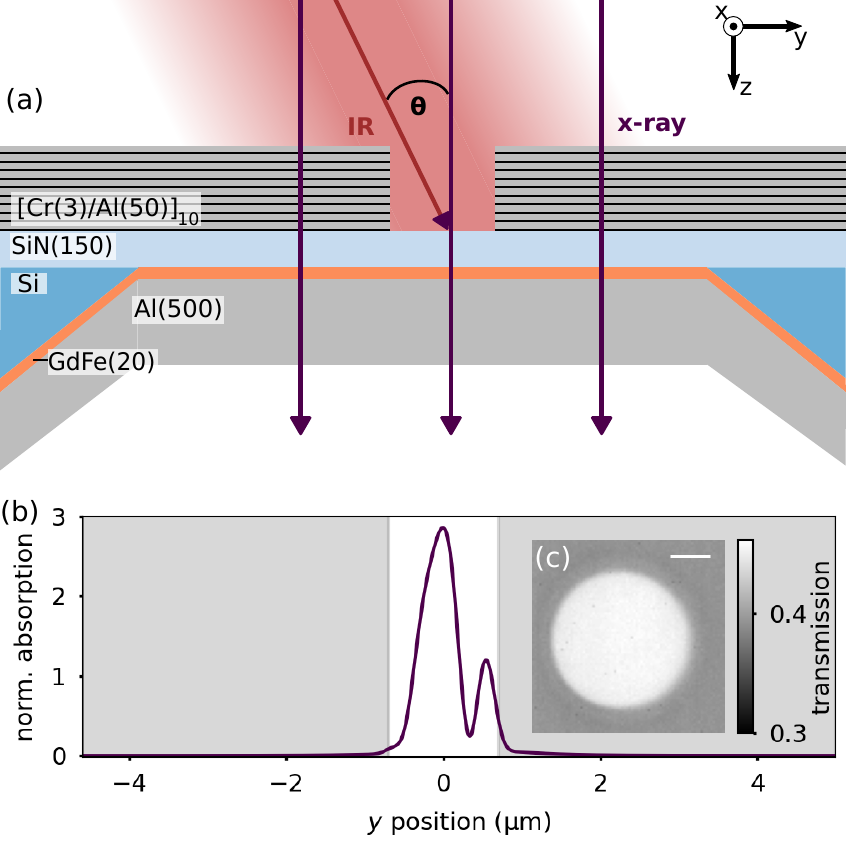}
    \caption{(a) Sample layout (vertical cross-section) for the time-resolved measurements (not to scale, layer thicknesses in nm).
    (b) Simulated vertical IR absorption profile within the $\mathrm{Gd}_{x}\mathrm{Fe}_{100-x}$ film after IR propagation through the aperture, incident under \SI{26}{\degree} with respect to the sample normal.
    The absorption was normalized to the absorption in the film without the Cr/Al mask.
    (c) Non-dichroic STXM image, $(I_+(x,y)  I_-(x,y))^{1/2}$, of the Cr/Al mask with aperture. Scalebar, \SI{500}{\nano\meter}.
    }
    \label{fig:sample}
\end{figure}

The ultrafast demagnetization process was induced via photo-excitation using a 1039-nm IR pump laser.
The laser oscillator is passively mode-locked with a laser-diode-pumped Yb:KGW crystal as active lasing element, operating at a tunable repetition rate of \SI{48.5}{\mega\hertz} to \SI{50.5}{\mega\hertz}.
After amplification in a Yb-doped fiber amplifier to a maximum pulse energy of \SI{50}{nJ} the repetition rate can be flexibly reduced down to single pulses via fast acousto-optic pulse selectors.
The laser is synchronized to the BESSY II bunch clock with an accuracy of \SI{0.3}{\pico\second} for time-resolved pump--probe experiments, which are carried out in standard multi-bunch operation of the storage ring.
The time resolution of these experiments is limited by the x-ray pulse duration of \SI{50}{\pico\second} (fwhm) which is not variable in the normal user operation mode of BESSY II.
Different pump--probe delays can be retrieved via an asynchronous excitation scheme,\cite{Bonetti2015, Weigand2022} which allows recording all time delays without the need for an optical delay stage.
However, as the initial phase between pump and probe is unknown, time-zero has to be determined from the dynamic effect observed, leading to a systematic error of $\pm 0.5$ time steps.
A \SI{8}{\meter} long fiber delivers the laser pulses from the optical laser table to the vacuum chamber of the STXM microscope.
At the microscope, the pulses are temporally compressed to any chosen duration between \SIrange[range-phrase={\ and }]{2}{30}{\pico\second}, with the minimum duration also depending on the pulse energy.
The pulses are then fed into a \SI{0.5}{\meter} long in-vacuum fiber that transports the pulses inside the STXM instrument.
All optical fibers preserve the polarization of the laser pulses.
A custom-developped micro-optics system focuses the pulses onto the sample with a spot size of \SI{6.5}{\micro \meter} (fwhm) via a micro-mirror attached to the mechanics of the order-sorting aperture.
The pulses thereby impinge under a nominal inclination angle of \SI{20 \pm 8}{\degree}
in the vertical direction, depending on the beam position on the micro-mirror.

The high pump--probe repetition rate of up to \SI{\approx 50}{MHz} poses particular challenges to the heat management in the sample. 
For example, a laser peak fluence already on the order of \SI{1}{\fluence}---as commonly used to excite magnetization dynamics---results in an average power density of \SI{50}{\kilo\watt\per\centi\meter\squared}, corresponding to an average power of \SI{24}{mW} incident on the sample within the micrometer-sized focal spot.
We took several measures to reduce the thermal load on the actual thin magnetic film and improve heat dissipation as sketched in Fig.~\ref{fig:sample}(a):
\begin{itemize}
    \item The whole sample is masked with a reflective $[\mathrm{Cr}(3)/\mathrm{Al}(50)]_{10}$ film (thicknesses in \si{\nano\meter}) including an aperture with \SI{1.4}{\micro\meter} diameter, fabricated via focused-ion beam (FIB) milling. 
    We geometrically estimate that the aperture selects only \SI{3.2}{\percent} of the IR beam.
    Otherwise, the Cr/Al mask reflects \SI{95}{\percent} of the incident IR radiation while maintaining a high transparency for the soft x-rays (see Fig.~\ref{fig:sample}(c)). 
    The remaining absorbed heat is dissipated within the mask layer itself due to thermal isolation by the silicon-nitride membrane. 
    This isolation prevents heat and hot-electron transport into the GdFe layer which otherwise would lead to an additional indirect excitation of demagnetization effects.\cite{Vodungbo_2016} 
    \item  
    The IR pulse first traverses the IR-transparent silicon-nitride substrate before hitting the magnetic layer, resulting in a refractive-index matching reducing the reflectivity of the layer structure.
    The magnetic film absorbs \SI{26}{\percent} of the incoming laser power compared to only \SI{10}{\percent} if the film was exposed to the IR directly, which helps reducing the overall pulse energy for photo-excitation.\cite{Kern2022-2}
    \item We use a \SI{500}{\nano\meter} thick Al heatsink layer directly deposited on the GdFe layer to effectively conduct the heat away from the magnetic film.
    \item 
    The STXM vacuum chamber is filled with He gas under a pressure of approximately \SI{10}{\milli\bar} to promote sample cooling.
\end{itemize}

The near-field diffraction at the edges of the circular aperture leads to an inhomogeneous excitation of the magnetic layer.\cite{vonKorffSchmising2014}
Finite-element simulations (COMSOL multiphysics) of the wave propagation predict two distinct maxima of the energy absorbed within the $\mathrm{Gd}\mathrm{Fe}$ layer inside the aperture as shown in Fig.~\ref{fig:sample}(b).
In this panel, the absorption is normalized to the absorption simulated without mask, showing an actual enhancement of a factor of up to almost \num{3} at the main peak and a pronounced quenching in between.
In the following, however, we refer to the laser fluence via the \emph{incident} peak fluence of the Gaussian laser focus.

We have performed heat-flow simulations to estimate the effectiveness of our measures to mitigate the heat load. As explained in detail in Appendix A, we found that, in particular, the combination of confining the excited area to micrometer size and applying a heatsink layer conducting the heat away from this area is highly suitable to ensure a moderate constant heating of the magnetic layer by the laser excitation during a time-resolved measurement. Using typical laser fluences to excite magnetization dynamics, the temperature rise of the layer during the measurement will stay below \SI{10}{K} even at \SI{50}{MHz} repetition rate. At the same time, the simulations show that the picosecond photo-excitation still allows transiently reaching high-temperature, out-of-equilibrium states in the sample which are inaccessible with slower means of excitation.

\section{Results and Discussion}

\begin{figure}
    \centering
    \includegraphics{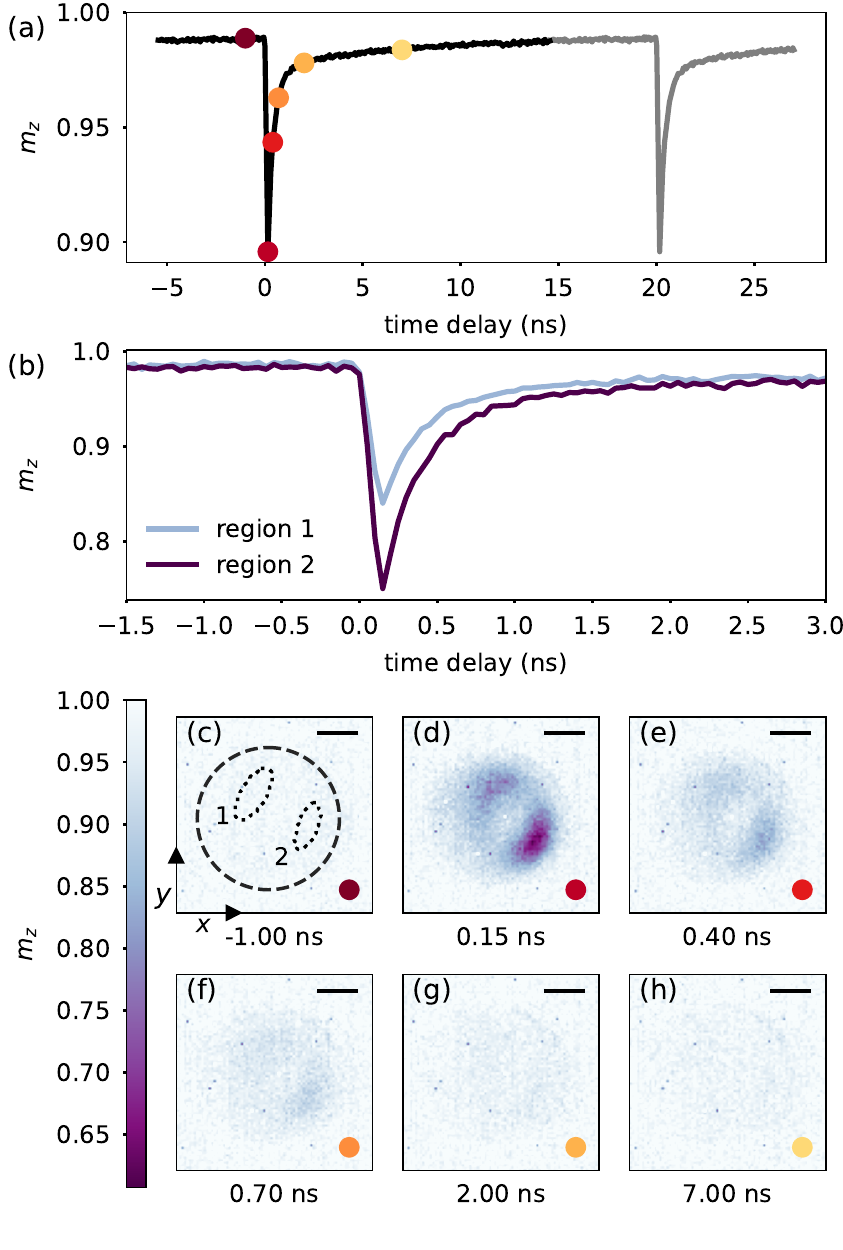}
    \caption{(a) Transient magnetization of Gd integrated over the size of the aperture in the Cr/Al mask in $\mathrm{Gd}_{29}\mathrm{Fe}_{71}$ after optical excitation at $t = 0$, measured with time steps of \SI{50}{\pico\second}. 
    The optical excitation has a repetition rate of \SI{50}{\mega\hertz}, the gray curve is a repetition of the black curve indicating the repetitive pump--probe cycle.  
    (b) Transient magnetization of Gd integrated over two separate regions as indicated in (c), coinciding with the field enhancement from the Cr/Al mask.
    (c)--(h) Magnetization of the sample at particular delay times as indicated in (a) by the colored dots. 
    The dashed circle in (c) shows the aperture in the Cr/Al mask.
    Scalebar, \SI{500}{\nano\metre}.
    }
    \label{fig:S2001i}
\end{figure}

We imaged the laser-induced de- and remagnetization of a $\mathrm{Gd}_{29}\mathrm{Fe}_{71}$ sample using a pump laser repetition rate of \SI{50}{\mega\hertz}, a pump pulse duration of \SI{4.4}{\pico\second} and an energy per pulse on the sample of \SI{1.5}{\nano\joule} corresponding to \SI{3.1}{\fluence} laser fluence. 
Exemplary XMCD images are shown in Fig.~\ref{fig:S2001i}(c)--(h) for different delays. 

The transient evolution of the magnetization integrated over the whole aperture (indicated by the dashed circle in Fig.~\ref{fig:S2001i}(c)) is shown in Fig.~\ref{fig:S2001i}(a). 
On average, the sample demagnetizes by \SI{\approx 10}{\percent}, rapidly reaching the magnetization minimum \SI{0.15}{\nano\second} after the photo-excitation. 
However, limited by the x-ray bunch duration of \SI{\approx 50}{\pico\second} (fwhm), we are certainly unable to resolve the dynamics of the Gd sublattice during the GdFe demagnetization, which is known from XMCD and TMOKE measurements with femtosecond temporal resolution to proceed on a sub-ps up to a few ps time scale.\cite{Radu2015, Hennecke2022}
After demagnetization, the magnetization recovers within \SI{2}{\nano\second} by \SI{90}{\percent}. 
We also observe that the magnetization level before time-zero is already slightly decreased by \SI{1.4}{\percent} with respect to the non-excited sample areas, which we attribute to a small static heating induced by the absorption of the pump pulses.
However, we do not observe any further indication of magnetic alteration or sample damage, which means that the sample withstood the photo-excitation at \SI{50}{\mega\hertz} repetition rate for several hours.

In addition to the well-known temporal evolution of the photo-excited demagnetization of GdFe, the STXM images provide nanometer-scale spatial information.
We indeed observe laterally very inhomogeneous demagnetization amplitudes (Fig.~\ref{fig:S2001i}(b)), which is particularly prominent at \SI{0.15}{\nano\second} after photo-excitation, where the sample demagnetizes locally by up to \SI{\approx 40}{\percent} (Fig.~\ref{fig:S2001i}(d)). 
This corresponds to the area where the intensity of the incoming laser pulse is amplified by the near-field diffraction induced by the aperture in the Cr/Al mask.\cite{vonKorffSchmising2014} 
The two prominent areas with increased demagnetization observed in the STXM images fit well to the double peak in the simulated near-field diffraction (Fig.~\ref{fig:sample}(b)) when assuming an incident angle of \SI{26}{\degree}. Furthermore we observe a tilt of the maxima with respect to the vertical direction which we attribute to an additional azimuthal tilt of the IR beam axis.

\begin{figure}
    \centering
    \includegraphics{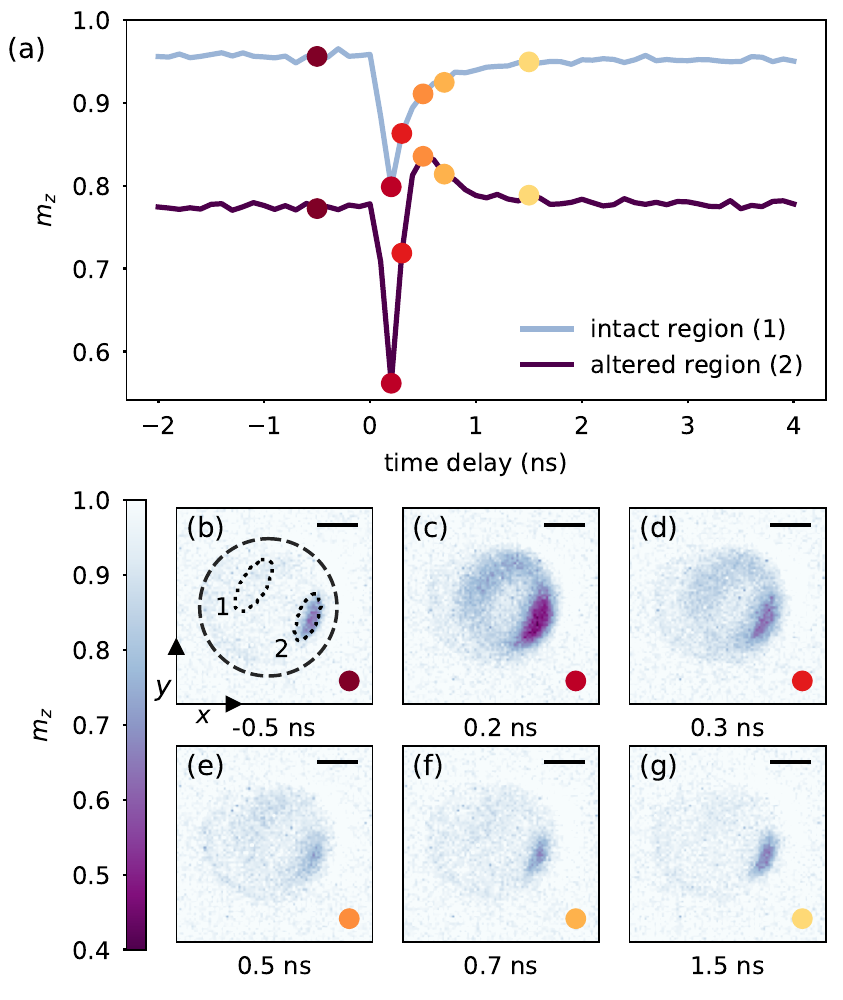}
    \caption{(a) Transient magnetization of Gd in $\mathrm{Gd}_{27}\mathrm{Fe}_{73}$ after optical excitation at $t = 0$, measured with time steps of \SI{100}{\pico \second}, and integrated over the magnetically intact and the magnetically altered regions as indicated in (b). 
    (b)--(g) Magnetization of the sample at particular delay times as indicated in (a) by the colored dots. 
    The dashed circle in (b) shows the aperture in the Cr/Al mask.
    Scalebar, \SI{500}{\nano\metre}.
    }
    \label{fig:S2001c}
\end{figure}

In order to investigate samples at even higher pump pulse energy, we lowered the repetition rate to \SI{5}{\mega\hertz} (\SI{18}{\pico\second} pulse duration) and performed measurements on the complementary $\mathrm{Gd}_{27}\mathrm{Fe}_{73}$ sample.
For a pump-pulse fluence of \SI{5.2}{\fluence} or below, we again do not observe any sample damage, resulting in magnetization dynamics very similar to Fig.~\ref{fig:S2001i} (not shown). However, further measurements at \SI{8.8}{\fluence} permanently altered the region of the sample exposed to the hotspot formed by the field enhancement inside the aperture.   
Here we show images of the sample from a time-resolved measurement (at \SI{9.8}{\fluence} IR fluence) recorded \emph{after} this modification was created (Fig.~\ref{fig:S2001c}(b)--(g)). 

The magnetically altered region is marked with a dotted line in Fig.~\ref{fig:S2001c}(b) (denoted as ``2'') and its magnetization transient is plotted in Fig.~\ref{fig:S2001c}(a). 
Compared to a magnetically intact region (dotted region ``1'' in Fig.~\ref{fig:S2001c}(b) and corresponding curve in Fig.~\ref{fig:S2001c}(a)), the response to the photo-excitation is remarkably different. 
Already the static magnetization (prior to photo-excitation) in the altered area is reduced by \SI{\approx 20}{\percent} in the out-of-plane direction with respect to the magnetization of the sample in regions which are still intact. 
Static XMCD data (not shown) taken after the time-resolved measurements confirm that this reduction is persistent and not induced by static heating during the pump--probe measurements.

After excitation, both the intact and the altered region demagnetize, while a larger demagnetization amplitude is observed in the altered region, which we mainly attribute to the higher laser intensity incident on this area. 
Interestingly, the magnetization in the altered region then recovers faster compared to the unaltered region and even transiently increases above the initial level. 
We observe this distinct dynamics in a confined region with a size of roughly \SIrange[range-phrase=$\times$, range-units=single]{600}{240}{\nano\meter\squared}.

In addition to its confined size, the average power used to create the modification (\SI{21}{\milli\watt}) is significantly smaller than the average power used to record the data in Fig.~\ref{fig:S2001i} (\SI{75}{\milli\watt}), where such a local material modification was not observed.
From these findings, we infer that the creation of the modification is induced by the energy of a single laser pulse and the corresponding highest temperature locally reached in the magnetic film rather than by static laser heating. 
It seems that the higher transient temperature reached in the second measurement leads to an annealing of the region in the magnetic alloy  where most of the IR is absorbed.

To explain the untypical magnetization dynamics, we speculate that the magnetization in the altered region is canted towards the in-plane direction due to a canted effective anisotropy. \cite{Chanda2021}
This magnetization canting would explain the lower out-of-plane magnetization before the pump pulse arrives. 
It is known, that rapid thermal annealing of GdFe alloys reduces the perpendicular magnetic anisotropy and for high temperatures even completely switches the magnetic easy axis from out-of-plane to in-plane.\cite{Talapatra2018, Talapatra2019}
When compared with unaltered regions, the remagnetization dynamics in the altered region provides some interesting insights on how the excited spin system is driven back to its initial state.
Apparently, the spins first follow the applied field while the anisotropy seems to be suppressed during this early, hot phase of the magnetization recovery up to \SI{\approx 0.5}{\nano\second}. The strong temperature dependence of the anisotropy is already known from static characterizations of GdFe alloys.\cite{Hansen1989}
Only at later times, when the film has further cooled down, the spins again align along the reestablished anisotropy axis. 
The ability to detect such distinct dynamics with a spatial resolution of \SI{30}{\nano\meter} showcases the potential of SR-based x-ray microscopy to study photo-induced magnetization phenomena.
Laser-based time-resolved x-ray microscopy provides access to the picosecond to nanosecond time scale where magnetic order typically reestablishes after photo-excitation.\cite{Buttner2021,Bergeard2014} This regime is of high relevance for technological applications as the speed of photo-induced magnetization switching is typically limited by the recovery process and nanometer-scale heat dissipation rather than the initial ultrafast transition to the excited state.\cite{Steinbach2022}

Our approach to manage the heat load on the sample largely builds on the reflection mask to confine the excitation to a micrometer-sized area and reduce the absorbed power in the actual sample. While we investigate a metallic thin-film sample, we consider this concept to be compatible with a variety of samples from materials science, chemistry and potentially biology which have been prepared for x-ray microscopy. The mask is highly transparent for soft x-rays (Fig.~\ref{fig:sample}(c)) leaving the area behind the mask observable with x-rays while being almost unaffected by the photo-excitation (Appendix A, Fig.~\ref{fig:appendix}(b)). However, the masking of the sample also results in an inhomogeneous excitation of the sample due to near-field diffraction, which requires customization and modeling of the mask layout. On the other hand, tailored near-field intensity distributions or plasmonic field enhancements may be integrated to provide a structured and localized optical excitation.\cite{LeGuyader2015, vonKorffSchmising2014, Yao2022, Kern2022-2, weder2020}

\section{Conclusions}

In summary, we demonstrated laser-pump--STXM-probe measurements with picosecond photo-excitation, using a new and now permanently installed fiber laser at the MAXYMUS instrument.
At incident IR laser fluences up to \SI{3}{\fluence}, we performed a destruction-free experiment on a magnetic thin-film sample with a repetition rate of \SI{50}{\mega\hertz}, studying laser-induced demagnetization and recovery in GdFe alloys. 
The repetition rate used here is orders of magnitude higher compared to what is typically used in experiments on photo-induced magnetization dynamics.\cite{Hennecke2022, Hassdenteufel2014, Moller2021}
With our approach, it is now possible to discern light-triggered picosecond dynamics with \SI{30}{\nano\meter} spatial resolution.
As many classes of materials are heterogeneous on a sub-micrometer length scale---be it intentionally or inadvertently, we expect this to be a very valuable capability to map out structural origins  of a material's functionality coupled to its dynamic behavior.
With a potential upgrade of the BESSY-II storage ring to the BESSY VSR concept (Variable pulse length Storage Ring)\cite{Jankowiak2015} combined with an upgrade of the laser at the MAXYMUS endstation, the time-resolution of the microscope could be improved to \SI{1}{\pico\second} in the future.
We also note that similar sample designs to mitigate heat load and sample damage as demonstrated in this work may also become important for experiments at x-ray free-electron lasers operating
at MHz repetition rates,\cite{hagstrom2022} allowing the combination of highest spatial and temporal resolution in the study of dynamic phenomena.

\section*{Acknowledgements}
Measurements were carried out at the MAXYMUS instrument at the BESSY II electron storage ring operated by the Helmholtz-Zentrum Berlin für Materialien und Energie. 
Financial support from the Leibniz Association via Grant No. K162/2018 (OptiSPIN) is acknowledged.

\section*{Appendix A: Simulation of the heat management in the sample}

\begin{figure}[bth!]
    \centering
    \includegraphics[width=\columnwidth]{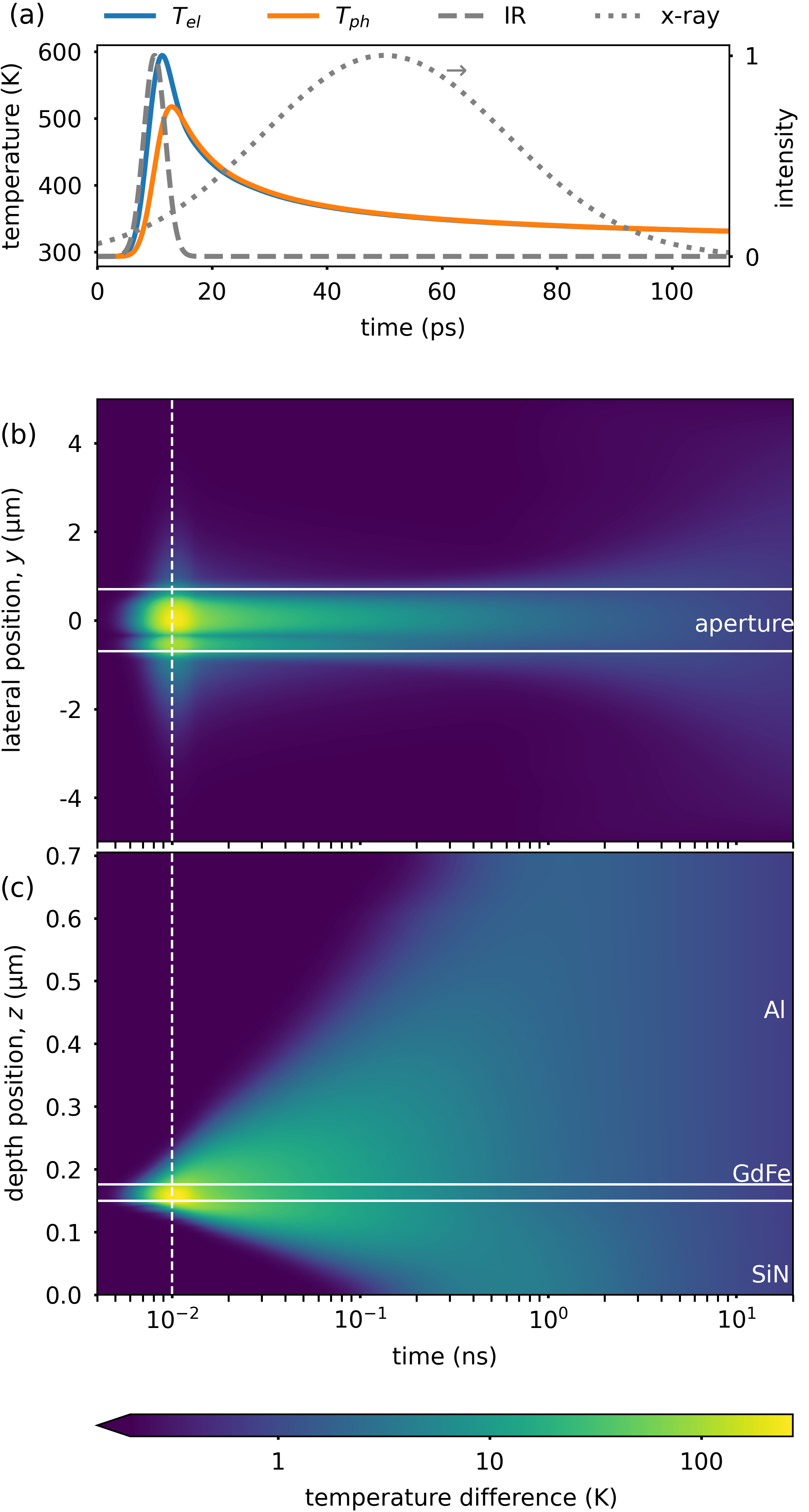}
    \caption{(a) Evolution of the electron temperature $T_\mathrm{el}$ and the phonon temperature $T_\mathrm{ph}$ in the GdFe layer after a laser excitation centered at \SI{10}{\pico\second} (gray dashed line).  The simulation includes heat transport along the depth of the sample. For comparison, the duration of the x-ray pulse is indicated as a Gaussian with a FWHM of \SI{50}{\pico\second} (dotted gray line).
    (b), (c) Evolution of increase of the phonon temperature (logarithmic pseudocolor scale) in the lateral direction within the GdFe layer and along the sample depth, respectively.
    The laser excitation is centered at \SI{10}{\pico\second} as indicated by the dashed white line.
    The solid white lines mark the position of the aperture (b) and the interfaces of the layers (c) as indicated.
    }
    \label{fig:appendix}
\end{figure}

The heat management during the time-resolved high-rep\-e\-ti\-tion-rate measurements is the most critical aspect of our experiments. We have performed a number of different simulations to assess the effectiveness of the measures applied in our experiments to mitigate the heat load on the sample from the excitation laser. In particular, the simulations shed light on the maximum temperature reached shortly after the laser pulse, and the time scale of the heat diffusion into the depth of the sample. Furthermore, we also simulate the lateral heat transport and the static heating resulting from the continuous high-repetition-rate excitation.

In the simulations, the magnetic sample is simplified to a stack of [Al(\SI{530}{\nano\meter})]/ Si$_3$N$_4$(\SI{150}{\nano\meter})/ GdFe(\SI{26}{\nano\meter})/ Al(\SI{530}{\nano\meter}), where the first Al layer is only present outside of the aperture. Material parameters used in the simulation for Al and Si$_3$N$_4$ can be found in Refs.~\citenum{Kern2022-2,weder2020}. For the GdFe layer, material parameters of the alloy were used if available\cite{Barker2013}; otherwise they were substituted by values of Fe.\cite{Johnson1974,Gray1972} We used the experimental parameters for the layout of the mask and the laser excitation at \SI{50}{\mega\hertz} repetition rate.

The response of the GdFe film to a single laser pulse at very early times (\SI{\le 1}{\nano\second}) can be simulated in a one-dimensional model neglecting the lateral heat transport, which drastically reduces the numerical effort. We have employed the udkm1Dsim package\cite{Schick2021} based on a two-temperature model (2TM) separating the response of electrons and phonons. The results are presented in Fig.~\ref{fig:appendix}(a) for the GdFe electron and phonon temperature. During the photo-excitation (centered at \SI{10}{\pico\second} in time), the temperatures increase almost instantaneously by about \SI{300}{\kelvin} and \SI{200}{\kelvin}, respectively. Soon after the maximum temperatures are reached, the temperatures of both systems equalize and the common temperature rapidly drops due to heat diffusion into the depth of the sample on a sub-100-ps time scale. As indicated by the (schematic) pulse shape of the x-ray detection, this very fast component of the dynamics can hardly be captured in our measurements in detail. However, we again note that the out-of-equilibrium high-temperature state after excitation may cause magnetization dynamics (also on longer time scales) that remains inaccessible with slower excitation.

In addition, we have performed two-dimensional simulations to also include the influence of the lateral heat transport and assess the impact of the Al heatsink layer. The simulations are carried out with the COMSOL multiphysics package again using a 2TM.\cite{Kern2022-2,weder2020} The geometry now includes the aperture in the top Al mask and has a total lateral width of \SI{10}{\micro\meter}. At the left and right borders, the system is coupled to a heat bath at room temperature. We show the results for the evolution and spatial dependence of the phonon temperature in the lateral dimension and the depth direction in Fig.~\ref{fig:appendix}(b) and (c), respectively. We again observe a picosecond high-temperature state during the photo-excitation and a very fast heat diffusion into the depth of the sample. This diffusion leads to an almost uniform temperature in the entire sample volume below the aperture after approximately \SI{1}{\nano\second}. Due to the larger dimensions the lateral diffusion is much slower. Nevertheless, within one pump--probe cycle of \SI{20}{\nano\second}, the heat has almost completely diffused from the excited volume into the whole sample. This finding underlines the effectiveness of the heatsink layer for our experiments.

In a last step, we further simplify the COMSOL model to simulate the long-term heat diffusion and temperature increase of the sample. The excitation is now simulated as continuous laser radiation with a power corresponding to the average absorbed power from our pulsed laser source inside the aperture. The simulation reveals that, on longer time scales (including many laser pulses), the temperature increases with a time constant of \SI{\sim 100}{\nano\second} and saturates at an increase of \SI{6}{\kelvin} above room temperature inside the primarily excited area of the GdFe film (not shown).

We compare this result with a simple analytical approximation that is particularly suitable for practical purposes. The heat-flow geometry in our sample can be perceived as heat conduction through a cylindrical wall. The inner part of the cylinder corresponds to the excited volume ($r_\mathrm{i} = \SI{0.7}{\micro\meter}$), the outer part corresponds to the macroscopic size of the entire sample which stays at constant temperature, e.g., the lateral size of the silicon-nitride membrane ($r_\mathrm{a} \approx \SI{100}{\micro\meter}$). The height of the cylindrical wall is given by the Al film (heatsink) thickness of $d = \SI{0.53}{\micro\meter}$. The resulting heat flow through the wall is:
\begin{eqnarray}
      P = 2 \pi d \lambda_\mathrm{c} \frac{\Delta T}{\ln(r_\mathrm{a}/r_\mathrm{i})}.
\end{eqnarray}
In equilibrium, this heat flow is balanced by the continuous heating from the laser excitation, which we (geometrically) estimate to an average absorbed power of $P = \SI{0.62}{\milli\watt}$ (see main text). Using the heat conductivity of Al of $\lambda_\mathrm{c} = \SI[per-mode = power]{239}{\watt\per\kelvin\per\meter}$ (Ref.~\citenum{Kuster1968}), we obtain a temperature increase inside the cylinder of $\Delta T \approx \SI{4}{\kelvin}$. Strictly speaking, this value corresponds to the temperature increase in the Al heatsink volume below the excited film area. However, as the GdFe layer is in direct contact to the heatsink, we expect only a slightly higher temperature in the GdFe which is in agreement with the result from the COMSOL simulation. We also note that $\Delta T$ only weakly depends on the precise choice of $r_\mathrm{a}$ and $r_\mathrm{i}$, making this estimate very robust. In conclusion, the very moderate temperature increase in the sample from the constant heating by the laser pulses as derived from numerical simulation and analytical approximation proves the performance of our concept of confining the excited film area to micrometer size in combination with using a heatsink layer to transport the heat away from this area.

\section*{Data Availability Statement}
The data that support the findings of this study are openly available in Zenodo at http://doi.org/10.5281/zenodo.7647267, reference number \citenum{Gerlinger2023}.

\bibliography{library}

%aipnum4-2.bst 2019-01-14 (MD) hand-edited version of apsrev4-1.bst
\begin{thebibliography}{45}%
\makeatletter
\providecommand \@ifxundefined [1]{%
 \@ifx{#1\undefined}
}%
\providecommand \@ifnum [1]{%
 \ifnum #1\expandafter \@firstoftwo
 \else \expandafter \@secondoftwo
 \fi
}%
\providecommand \@ifx [1]{%
 \ifx #1\expandafter \@firstoftwo
 \else \expandafter \@secondoftwo
 \fi
}%
\providecommand \natexlab [1]{#1}%
\providecommand \enquote  [1]{``#1''}%
\providecommand \bibnamefont  [1]{#1}%
\providecommand \bibfnamefont [1]{#1}%
\providecommand \citenamefont [1]{#1}%
\providecommand \href@noop [0]{\@secondoftwo}%
\providecommand \href [0]{\begingroup \@sanitize@url \@href}%
\providecommand \@href[1]{\@@startlink{#1}\@@href}%
\providecommand \@@href[1]{\endgroup#1\@@endlink}%
\providecommand \@sanitize@url [0]{\catcode `\\12\catcode `\$12\catcode
  `\&12\catcode `\#12\catcode `\^12\catcode `\_12\catcode `\%12\relax}%
\providecommand \@@startlink[1]{}%
\providecommand \@@endlink[0]{}%
\providecommand \url  [0]{\begingroup\@sanitize@url \@url }%
\providecommand \@url [1]{\endgroup\@href {#1}{\urlprefix }}%
\providecommand \urlprefix  [0]{URL }%
\providecommand \Eprint [0]{\href }%
\providecommand \doibase [0]{https://doi.org/}%
\providecommand \selectlanguage [0]{\@gobble}%
\providecommand \bibinfo  [0]{\@secondoftwo}%
\providecommand \bibfield  [0]{\@secondoftwo}%
\providecommand \translation [1]{[#1]}%
\providecommand \BibitemOpen [0]{}%
\providecommand \bibitemStop [0]{}%
\providecommand \bibitemNoStop [0]{.\EOS\space}%
\providecommand \EOS [0]{\spacefactor3000\relax}%
\providecommand \BibitemShut  [1]{\csname bibitem#1\endcsname}%
\let\auto@bib@innerbib\@empty
%</preamble>
\bibitem [{\citenamefont {Vedmedenko}\ \emph {et~al.}(2020)\citenamefont
  {Vedmedenko}, \citenamefont {Kawakami}, \citenamefont {Sheka}, \citenamefont
  {Gambardella}, \citenamefont {Kirilyuk}, \citenamefont {Hirohata},
  \citenamefont {Binek}, \citenamefont {Chubykalo-Fesenko}, \citenamefont
  {Sanvito}, \citenamefont {Kirby}, \citenamefont {Grollier}, \citenamefont
  {Everschor-Sitte}, \citenamefont {Kampfrath}, \citenamefont {You},\ and\
  \citenamefont {Berger}}]{Vedmedenko2020}%
  \BibitemOpen
  \bibfield  {author} {\bibinfo {author} {\bibfnamefont {E.~Y.}\ \bibnamefont
  {Vedmedenko}}, \bibinfo {author} {\bibfnamefont {R.~K.}\ \bibnamefont
  {Kawakami}}, \bibinfo {author} {\bibfnamefont {D.~D.}\ \bibnamefont {Sheka}},
  \bibinfo {author} {\bibfnamefont {P.}~\bibnamefont {Gambardella}}, \bibinfo
  {author} {\bibfnamefont {A.}~\bibnamefont {Kirilyuk}}, \bibinfo {author}
  {\bibfnamefont {A.}~\bibnamefont {Hirohata}}, \bibinfo {author}
  {\bibfnamefont {C.}~\bibnamefont {Binek}}, \bibinfo {author} {\bibfnamefont
  {O.}~\bibnamefont {Chubykalo-Fesenko}}, \bibinfo {author} {\bibfnamefont
  {S.}~\bibnamefont {Sanvito}}, \bibinfo {author} {\bibfnamefont {B.~J.}\
  \bibnamefont {Kirby}}, \bibinfo {author} {\bibfnamefont {J.}~\bibnamefont
  {Grollier}}, \bibinfo {author} {\bibfnamefont {K.}~\bibnamefont
  {Everschor-Sitte}}, \bibinfo {author} {\bibfnamefont {T.}~\bibnamefont
  {Kampfrath}}, \bibinfo {author} {\bibfnamefont {C.-Y.}\ \bibnamefont {You}},\
  and\ \bibinfo {author} {\bibfnamefont {A.}~\bibnamefont {Berger}},\
  }\bibfield  {title} {\enquote {\bibinfo {title} {The 2020 magnetism
  roadmap},}\ }\href {https://doi.org/10.1088/1361-6463/ab9d98} {\bibfield
  {journal} {\bibinfo  {journal} {Journal of Physics D: Applied Physics}\
  }\textbf {\bibinfo {volume} {53}},\ \bibinfo {pages} {453001} (\bibinfo
  {year} {2020})}\BibitemShut {NoStop}%
\bibitem [{\citenamefont {Kent}\ \emph {et~al.}(2020)\citenamefont {Kent},
  \citenamefont {Ohldag}, \citenamefont {D{\"u}rr},\ and\ \citenamefont
  {Sun}}]{kent2020}%
  \BibitemOpen
  \bibfield  {author} {\bibinfo {author} {\bibfnamefont {A.~D.}\ \bibnamefont
  {Kent}}, \bibinfo {author} {\bibfnamefont {H.}~\bibnamefont {Ohldag}},
  \bibinfo {author} {\bibfnamefont {H.~A.}\ \bibnamefont {D{\"u}rr}},\ and\
  \bibinfo {author} {\bibfnamefont {J.~Z.}\ \bibnamefont {Sun}},\ }\bibfield
  {title} {\enquote {\bibinfo {title} {Magnetization {{Dynamics}}},}\ }in\
  \href {https://doi.org/10.1007/978-3-030-63101-7_27-1} {\emph {\bibinfo
  {booktitle} {Handbook of {{Magnetism}} and {{Magnetic Materials}}}}},\
  \bibinfo {editor} {edited by\ \bibinfo {editor} {\bibfnamefont
  {M.}~\bibnamefont {Coey}}\ and\ \bibinfo {editor} {\bibfnamefont
  {S.}~\bibnamefont {Parkin}}}\ (\bibinfo  {publisher} {{Springer International
  Publishing}},\ \bibinfo {address} {{Cham}},\ \bibinfo {year} {2020})\ pp.\
  \bibinfo {pages} {1--33}\BibitemShut {NoStop}%
\bibitem [{\citenamefont {Stoll}\ \emph {et~al.}(2015)\citenamefont {Stoll},
  \citenamefont {Noske}, \citenamefont {Weigand}, \citenamefont {Richter},
  \citenamefont {Kr{\"u}ger}, \citenamefont {Reeve}, \citenamefont {H{\"a}nze},
  \citenamefont {Adolff}, \citenamefont {Stein}, \citenamefont {Meier},
  \citenamefont {Kl{\"a}ui},\ and\ \citenamefont {Sch{\"u}tz}}]{stoll2015}%
  \BibitemOpen
  \bibfield  {author} {\bibinfo {author} {\bibfnamefont {H.}~\bibnamefont
  {Stoll}}, \bibinfo {author} {\bibfnamefont {M.}~\bibnamefont {Noske}},
  \bibinfo {author} {\bibfnamefont {M.}~\bibnamefont {Weigand}}, \bibinfo
  {author} {\bibfnamefont {K.}~\bibnamefont {Richter}}, \bibinfo {author}
  {\bibfnamefont {B.}~\bibnamefont {Kr{\"u}ger}}, \bibinfo {author}
  {\bibfnamefont {R.~M.}\ \bibnamefont {Reeve}}, \bibinfo {author}
  {\bibfnamefont {M.}~\bibnamefont {H{\"a}nze}}, \bibinfo {author}
  {\bibfnamefont {C.~F.}\ \bibnamefont {Adolff}}, \bibinfo {author}
  {\bibfnamefont {F.-U.}\ \bibnamefont {Stein}}, \bibinfo {author}
  {\bibfnamefont {G.}~\bibnamefont {Meier}}, \bibinfo {author} {\bibfnamefont
  {M.}~\bibnamefont {Kl{\"a}ui}},\ and\ \bibinfo {author} {\bibfnamefont
  {G.}~\bibnamefont {Sch{\"u}tz}},\ }\bibfield  {title} {\enquote {\bibinfo
  {title} {Imaging spin dynamics on the nanoscale using {{X-Ray}}
  microscopy},}\ }\href@noop {} {\bibfield  {journal} {\bibinfo  {journal}
  {Frontiers in Physics}\ }\textbf {\bibinfo {volume} {3}} (\bibinfo {year}
  {2015})}\BibitemShut {NoStop}%
\bibitem [{\citenamefont {Fischer}(2017)}]{fischer2017}%
  \BibitemOpen
  \bibfield  {author} {\bibinfo {author} {\bibfnamefont {P.}~\bibnamefont
  {Fischer}},\ }\bibfield  {title} {\enquote {\bibinfo {title} {Magnetic
  imaging with polarized soft x-rays},}\ }\href
  {https://doi.org/10.1088/1361-6463/aa778a} {\bibfield  {journal} {\bibinfo
  {journal} {Journal of Physics D: Applied Physics}\ }\textbf {\bibinfo
  {volume} {50}},\ \bibinfo {pages} {313002} (\bibinfo {year}
  {2017})}\BibitemShut {NoStop}%
\bibitem [{\citenamefont {Kirilyuk}, \citenamefont {Kimel},\ and\ \citenamefont
  {Rasing}(2010)}]{Kirilyuk2010}%
  \BibitemOpen
  \bibfield  {author} {\bibinfo {author} {\bibfnamefont {A.}~\bibnamefont
  {Kirilyuk}}, \bibinfo {author} {\bibfnamefont {A.~V.}\ \bibnamefont
  {Kimel}},\ and\ \bibinfo {author} {\bibfnamefont {T.}~\bibnamefont
  {Rasing}},\ }\bibfield  {title} {\enquote {\bibinfo {title} {Ultrafast
  optical manipulation of magnetic order},}\ }\href
  {https://doi.org/10.1103/RevModPhys.82.2731} {\bibfield  {journal} {\bibinfo
  {journal} {Rev. Mod. Phys.}\ }\textbf {\bibinfo {volume} {82}},\ \bibinfo
  {pages} {2731--2784} (\bibinfo {year} {2010})}\BibitemShut {NoStop}%
\bibitem [{\citenamefont {Carva}, \citenamefont {Bal{\'a}{\v{z}}},\ and\
  \citenamefont {Radu}(2017)}]{Carva2017}%
  \BibitemOpen
  \bibfield  {author} {\bibinfo {author} {\bibfnamefont {K.}~\bibnamefont
  {Carva}}, \bibinfo {author} {\bibfnamefont {P.}~\bibnamefont
  {Bal{\'a}{\v{z}}}},\ and\ \bibinfo {author} {\bibfnamefont {I.}~\bibnamefont
  {Radu}},\ }\href {https://doi.org/https://doi.org/10.1016/bs.hmm.2017.09.003}
  {\emph {\bibinfo {title} {Chapter 2 - Laser-Induced Ultrafast Magnetic
  Phenomena}}},\ edited by\ \bibinfo {editor} {\bibfnamefont {E.}~\bibnamefont
  {Br{\"u}ck}},\ \bibinfo {series} {Handbook of Magnetic Materials},
  Vol.~\bibinfo {volume} {26}\ (\bibinfo  {publisher} {Elsevier},\ \bibinfo
  {year} {2017})\ pp.\ \bibinfo {pages} {291--463}\BibitemShut {NoStop}%
\bibitem [{\citenamefont {Büttner}\ \emph {et~al.}(2021)\citenamefont
  {Büttner}, \citenamefont {Pfau}, \citenamefont {Böttcher}, \citenamefont
  {Schneider}, \citenamefont {Mercurio}, \citenamefont {Günther},
  \citenamefont {Hessing}, \citenamefont {Klose}, \citenamefont {Wittmann},
  \citenamefont {Gerlinger}, \citenamefont {Kern}, \citenamefont {Strüber},
  \citenamefont {von Korff~Schmising}, \citenamefont {Fuchs}, \citenamefont
  {Engel}, \citenamefont {Churikova}, \citenamefont {Huang}, \citenamefont
  {Suzuki}, \citenamefont {Lemesh}, \citenamefont {Huang}, \citenamefont
  {Caretta}, \citenamefont {Weder}, \citenamefont {Gaida}, \citenamefont
  {Möller}, \citenamefont {Harvey}, \citenamefont {Zayko}, \citenamefont
  {Bagschik}, \citenamefont {Carley}, \citenamefont {Mercadier}, \citenamefont
  {Schlappa}, \citenamefont {Yaroslavtsev}, \citenamefont {Le~Guyarder},
  \citenamefont {Gerasimova}, \citenamefont {Scherz}, \citenamefont {Deiter},
  \citenamefont {Gort}, \citenamefont {Hickin}, \citenamefont {Zhu},
  \citenamefont {Turcato}, \citenamefont {Lomidze}, \citenamefont {Erdinger},
  \citenamefont {Castoldi}, \citenamefont {Maffessanti}, \citenamefont {Porro},
  \citenamefont {Samartsev}, \citenamefont {Sinova}, \citenamefont {Ropers},
  \citenamefont {Mentink}, \citenamefont {Dupé}, \citenamefont {Beach},\ and\
  \citenamefont {Eisebitt}}]{Buttner2021}%
  \BibitemOpen
  \bibfield  {author} {\bibinfo {author} {\bibfnamefont {F.}~\bibnamefont
  {Büttner}}, \bibinfo {author} {\bibfnamefont {B.}~\bibnamefont {Pfau}},
  \bibinfo {author} {\bibfnamefont {M.}~\bibnamefont {Böttcher}}, \bibinfo
  {author} {\bibfnamefont {M.}~\bibnamefont {Schneider}}, \bibinfo {author}
  {\bibfnamefont {G.}~\bibnamefont {Mercurio}}, \bibinfo {author}
  {\bibfnamefont {C.~M.}\ \bibnamefont {Günther}}, \bibinfo {author}
  {\bibfnamefont {P.}~\bibnamefont {Hessing}}, \bibinfo {author} {\bibfnamefont
  {C.}~\bibnamefont {Klose}}, \bibinfo {author} {\bibfnamefont
  {A.}~\bibnamefont {Wittmann}}, \bibinfo {author} {\bibfnamefont
  {K.}~\bibnamefont {Gerlinger}}, \bibinfo {author} {\bibfnamefont {L.-M.}\
  \bibnamefont {Kern}}, \bibinfo {author} {\bibfnamefont {C.}~\bibnamefont
  {Strüber}}, \bibinfo {author} {\bibfnamefont {C.}~\bibnamefont {von
  Korff~Schmising}}, \bibinfo {author} {\bibfnamefont {J.}~\bibnamefont
  {Fuchs}}, \bibinfo {author} {\bibfnamefont {D.}~\bibnamefont {Engel}},
  \bibinfo {author} {\bibfnamefont {A.}~\bibnamefont {Churikova}}, \bibinfo
  {author} {\bibfnamefont {S.}~\bibnamefont {Huang}}, \bibinfo {author}
  {\bibfnamefont {D.}~\bibnamefont {Suzuki}}, \bibinfo {author} {\bibfnamefont
  {I.}~\bibnamefont {Lemesh}}, \bibinfo {author} {\bibfnamefont
  {M.}~\bibnamefont {Huang}}, \bibinfo {author} {\bibfnamefont
  {L.}~\bibnamefont {Caretta}}, \bibinfo {author} {\bibfnamefont
  {D.}~\bibnamefont {Weder}}, \bibinfo {author} {\bibfnamefont {J.~H.}\
  \bibnamefont {Gaida}}, \bibinfo {author} {\bibfnamefont {M.}~\bibnamefont
  {Möller}}, \bibinfo {author} {\bibfnamefont {T.~R.}\ \bibnamefont {Harvey}},
  \bibinfo {author} {\bibfnamefont {S.}~\bibnamefont {Zayko}}, \bibinfo
  {author} {\bibfnamefont {K.}~\bibnamefont {Bagschik}}, \bibinfo {author}
  {\bibfnamefont {R.}~\bibnamefont {Carley}}, \bibinfo {author} {\bibfnamefont
  {L.}~\bibnamefont {Mercadier}}, \bibinfo {author} {\bibfnamefont
  {J.}~\bibnamefont {Schlappa}}, \bibinfo {author} {\bibfnamefont
  {A.}~\bibnamefont {Yaroslavtsev}}, \bibinfo {author} {\bibfnamefont
  {L.}~\bibnamefont {Le~Guyarder}}, \bibinfo {author} {\bibfnamefont
  {N.}~\bibnamefont {Gerasimova}}, \bibinfo {author} {\bibfnamefont
  {A.}~\bibnamefont {Scherz}}, \bibinfo {author} {\bibfnamefont
  {C.}~\bibnamefont {Deiter}}, \bibinfo {author} {\bibfnamefont
  {R.}~\bibnamefont {Gort}}, \bibinfo {author} {\bibfnamefont {D.}~\bibnamefont
  {Hickin}}, \bibinfo {author} {\bibfnamefont {J.}~\bibnamefont {Zhu}},
  \bibinfo {author} {\bibfnamefont {M.}~\bibnamefont {Turcato}}, \bibinfo
  {author} {\bibfnamefont {D.}~\bibnamefont {Lomidze}}, \bibinfo {author}
  {\bibfnamefont {F.}~\bibnamefont {Erdinger}}, \bibinfo {author}
  {\bibfnamefont {A.}~\bibnamefont {Castoldi}}, \bibinfo {author}
  {\bibfnamefont {S.}~\bibnamefont {Maffessanti}}, \bibinfo {author}
  {\bibfnamefont {M.}~\bibnamefont {Porro}}, \bibinfo {author} {\bibfnamefont
  {A.}~\bibnamefont {Samartsev}}, \bibinfo {author} {\bibfnamefont
  {J.}~\bibnamefont {Sinova}}, \bibinfo {author} {\bibfnamefont
  {C.}~\bibnamefont {Ropers}}, \bibinfo {author} {\bibfnamefont {J.~H.}\
  \bibnamefont {Mentink}}, \bibinfo {author} {\bibfnamefont {B.}~\bibnamefont
  {Dupé}}, \bibinfo {author} {\bibfnamefont {G.~S.~D.}\ \bibnamefont
  {Beach}},\ and\ \bibinfo {author} {\bibfnamefont {S.}~\bibnamefont
  {Eisebitt}},\ }\bibfield  {title} {\enquote {\bibinfo {title} {Observation of
  fluctuation-mediated picosecond nucleation of a topological phase},}\ }\href
  {https://doi.org/10.1038/s41563-020-00807-1} {\bibfield  {journal} {\bibinfo
  {journal} {Nature Materials}\ }\textbf {\bibinfo {volume} {20}},\ \bibinfo
  {pages} {30--37} (\bibinfo {year} {2021})}\BibitemShut {NoStop}%
\bibitem [{\citenamefont {Gerlinger}\ \emph {et~al.}(2021)\citenamefont
  {Gerlinger}, \citenamefont {Pfau}, \citenamefont {Büttner}, \citenamefont
  {Schneider}, \citenamefont {Kern}, \citenamefont {Fuchs}, \citenamefont
  {Engel}, \citenamefont {Günther}, \citenamefont {Huang}, \citenamefont
  {Lemesh}, \citenamefont {Caretta}, \citenamefont {Churikova}, \citenamefont
  {Hessing}, \citenamefont {Klose}, \citenamefont {Strüber}, \citenamefont
  {Litzius}, \citenamefont {Metternich}, \citenamefont {Battistelli},
  \citenamefont {Bagschik}, \citenamefont {Sadovnikov},\ and\ \citenamefont
  {Eisebitt}}]{Gerlinger2021}%
  \BibitemOpen
  \bibfield  {author} {\bibinfo {author} {\bibfnamefont {K.}~\bibnamefont
  {Gerlinger}}, \bibinfo {author} {\bibfnamefont {B.}~\bibnamefont {Pfau}},
  \bibinfo {author} {\bibfnamefont {F.}~\bibnamefont {Büttner}}, \bibinfo
  {author} {\bibfnamefont {M.}~\bibnamefont {Schneider}}, \bibinfo {author}
  {\bibfnamefont {L.-M.}\ \bibnamefont {Kern}}, \bibinfo {author}
  {\bibfnamefont {J.}~\bibnamefont {Fuchs}}, \bibinfo {author} {\bibfnamefont
  {D.}~\bibnamefont {Engel}}, \bibinfo {author} {\bibfnamefont {C.~M.}\
  \bibnamefont {Günther}}, \bibinfo {author} {\bibfnamefont {M.}~\bibnamefont
  {Huang}}, \bibinfo {author} {\bibfnamefont {I.}~\bibnamefont {Lemesh}},
  \bibinfo {author} {\bibfnamefont {L.}~\bibnamefont {Caretta}}, \bibinfo
  {author} {\bibfnamefont {A.}~\bibnamefont {Churikova}}, \bibinfo {author}
  {\bibfnamefont {P.}~\bibnamefont {Hessing}}, \bibinfo {author} {\bibfnamefont
  {C.}~\bibnamefont {Klose}}, \bibinfo {author} {\bibfnamefont
  {C.}~\bibnamefont {Strüber}}, \bibinfo {author} {\bibfnamefont
  {K.}~\bibnamefont {Litzius}}, \bibinfo {author} {\bibfnamefont
  {D.}~\bibnamefont {Metternich}}, \bibinfo {author} {\bibfnamefont
  {R.}~\bibnamefont {Battistelli}}, \bibinfo {author} {\bibfnamefont
  {K.}~\bibnamefont {Bagschik}}, \bibinfo {author} {\bibfnamefont
  {A.}~\bibnamefont {Sadovnikov}},\ and\ \bibinfo {author} {\bibfnamefont
  {S.}~\bibnamefont {Eisebitt}},\ }\bibfield  {title} {\enquote {\bibinfo
  {title} {Application concepts for ultrafast laser-induced skyrmion creation
  and annihilation},}\ }\href@noop {} {\bibfield  {journal} {\bibinfo
  {journal} {Applied Physics Letters}\ }\textbf {\bibinfo {volume} {118}},\
  \bibinfo {pages} {7} (\bibinfo {year} {2021})}\BibitemShut {NoStop}%
\bibitem [{\citenamefont {Subkhangulov}\ \emph {et~al.}(2015)\citenamefont
  {Subkhangulov}, \citenamefont {Henriques}, \citenamefont {Rappl},
  \citenamefont {Abramof}, \citenamefont {Rasing},\ and\ \citenamefont
  {Kimel}}]{Subkhangulov2015}%
  \BibitemOpen
  \bibfield  {author} {\bibinfo {author} {\bibfnamefont {R.~R.}\ \bibnamefont
  {Subkhangulov}}, \bibinfo {author} {\bibfnamefont {A.~B.}\ \bibnamefont
  {Henriques}}, \bibinfo {author} {\bibfnamefont {P.~H.~O.}\ \bibnamefont
  {Rappl}}, \bibinfo {author} {\bibfnamefont {E.}~\bibnamefont {Abramof}},
  \bibinfo {author} {\bibfnamefont {T.}~\bibnamefont {Rasing}},\ and\ \bibinfo
  {author} {\bibfnamefont {A.~V.}\ \bibnamefont {Kimel}},\ }\bibfield  {title}
  {\enquote {\bibinfo {title} {All-optical manipulation and probing of the
  d–f exchange interaction in {EuTe}},}\ }\href
  {https://doi.org/10.1038/srep04368} {\bibfield  {journal} {\bibinfo
  {journal} {Scientific Reports}\ }\textbf {\bibinfo {volume} {4}},\ \bibinfo
  {pages} {4368} (\bibinfo {year} {2015})}\BibitemShut {NoStop}%
\bibitem [{\citenamefont {Satoh}\ \emph {et~al.}(2012)\citenamefont {Satoh},
  \citenamefont {Terui}, \citenamefont {Moriya}, \citenamefont {Ivanov},
  \citenamefont {Ando}, \citenamefont {Saitoh}, \citenamefont {Shimura},\ and\
  \citenamefont {Kuroda}}]{Satoh2012}%
  \BibitemOpen
  \bibfield  {author} {\bibinfo {author} {\bibfnamefont {T.}~\bibnamefont
  {Satoh}}, \bibinfo {author} {\bibfnamefont {Y.}~\bibnamefont {Terui}},
  \bibinfo {author} {\bibfnamefont {R.}~\bibnamefont {Moriya}}, \bibinfo
  {author} {\bibfnamefont {B.~A.}\ \bibnamefont {Ivanov}}, \bibinfo {author}
  {\bibfnamefont {K.}~\bibnamefont {Ando}}, \bibinfo {author} {\bibfnamefont
  {E.}~\bibnamefont {Saitoh}}, \bibinfo {author} {\bibfnamefont
  {T.}~\bibnamefont {Shimura}},\ and\ \bibinfo {author} {\bibfnamefont
  {K.}~\bibnamefont {Kuroda}},\ }\bibfield  {title} {\enquote {\bibinfo {title}
  {Directional control of spin-wave emission by spatially shaped light},}\
  }\href {https://doi.org/10.1038/nphoton.2012.218} {\bibfield  {journal}
  {\bibinfo  {journal} {Nature Photonics}\ }\textbf {\bibinfo {volume} {6}},\
  \bibinfo {pages} {662--666} (\bibinfo {year} {2012})}\BibitemShut {NoStop}%
\bibitem [{\citenamefont {Schellekens}\ \emph {et~al.}(2014)\citenamefont
  {Schellekens}, \citenamefont {Kuiper}, \citenamefont {de~Wit},\ and\
  \citenamefont {Koopmans}}]{Schellekens2014}%
  \BibitemOpen
  \bibfield  {author} {\bibinfo {author} {\bibfnamefont {A.~J.}\ \bibnamefont
  {Schellekens}}, \bibinfo {author} {\bibfnamefont {K.~C.}\ \bibnamefont
  {Kuiper}}, \bibinfo {author} {\bibfnamefont {R.~R. J.~C.}\ \bibnamefont
  {de~Wit}},\ and\ \bibinfo {author} {\bibfnamefont {B.}~\bibnamefont
  {Koopmans}},\ }\bibfield  {title} {\enquote {\bibinfo {title} {Ultrafast
  spin-transfer torque driven by femtosecond pulsed-laser excitation},}\ }\href
  {https://doi.org/10.1038/ncomms5333} {\bibfield  {journal} {\bibinfo
  {journal} {Nature Communications}\ }\textbf {\bibinfo {volume} {5}},\
  \bibinfo {pages} {4333} (\bibinfo {year} {2014})}\BibitemShut {NoStop}%
\bibitem [{\citenamefont {Kim}, \citenamefont {Vomir},\ and\ \citenamefont
  {Bigot}(2015)}]{Kim2015}%
  \BibitemOpen
  \bibfield  {author} {\bibinfo {author} {\bibfnamefont {J.-W.}\ \bibnamefont
  {Kim}}, \bibinfo {author} {\bibfnamefont {M.}~\bibnamefont {Vomir}},\ and\
  \bibinfo {author} {\bibfnamefont {J.-Y.}\ \bibnamefont {Bigot}},\ }\bibfield
  {title} {\enquote {\bibinfo {title} {Controlling the {Spins} {Angular}
  {Momentum} in {Ferromagnets} with {Sequences} of {Picosecond} {Acoustic}
  {Pulses}},}\ }\href {https://doi.org/10.1038/srep08511} {\bibfield  {journal}
  {\bibinfo  {journal} {Scientific Reports}\ }\textbf {\bibinfo {volume} {5}},\
  \bibinfo {pages} {8511} (\bibinfo {year} {2015})}\BibitemShut {NoStop}%
\bibitem [{\citenamefont {Choe}\ \emph {et~al.}(2004)\citenamefont {Choe},
  \citenamefont {Acremann}, \citenamefont {Scholl}, \citenamefont {Bauer},
  \citenamefont {Doran}, \citenamefont {Stöhr},\ and\ \citenamefont
  {Padmore}}]{Choe2004}%
  \BibitemOpen
  \bibfield  {author} {\bibinfo {author} {\bibfnamefont {S.-B.}\ \bibnamefont
  {Choe}}, \bibinfo {author} {\bibfnamefont {Y.}~\bibnamefont {Acremann}},
  \bibinfo {author} {\bibfnamefont {A.}~\bibnamefont {Scholl}}, \bibinfo
  {author} {\bibfnamefont {A.}~\bibnamefont {Bauer}}, \bibinfo {author}
  {\bibfnamefont {A.}~\bibnamefont {Doran}}, \bibinfo {author} {\bibfnamefont
  {J.}~\bibnamefont {Stöhr}},\ and\ \bibinfo {author} {\bibfnamefont {H.~A.}\
  \bibnamefont {Padmore}},\ }\bibfield  {title} {\enquote {\bibinfo {title}
  {Vortex {Core}-{Driven} {Magnetization} {Dynamics}},}\ }\href
  {https://doi.org/10.1126/science.1095068} {\bibfield  {journal} {\bibinfo
  {journal} {Science}\ }\textbf {\bibinfo {volume} {304}},\ \bibinfo {pages}
  {420--422} (\bibinfo {year} {2004})}\BibitemShut {NoStop}%
\bibitem [{\citenamefont {Heyne}\ \emph {et~al.}(2010)\citenamefont {Heyne},
  \citenamefont {Kläui}, \citenamefont {Rhensius}, \citenamefont
  {Le~Guyader},\ and\ \citenamefont {Nolting}}]{Heyne2010}%
  \BibitemOpen
  \bibfield  {author} {\bibinfo {author} {\bibfnamefont {L.}~\bibnamefont
  {Heyne}}, \bibinfo {author} {\bibfnamefont {M.}~\bibnamefont {Kläui}},
  \bibinfo {author} {\bibfnamefont {J.}~\bibnamefont {Rhensius}}, \bibinfo
  {author} {\bibfnamefont {L.}~\bibnamefont {Le~Guyader}},\ and\ \bibinfo
  {author} {\bibfnamefont {F.}~\bibnamefont {Nolting}},\ }\bibfield  {title}
  {\enquote {\bibinfo {title} {\textit{{In} situ} contacting and
  current-injection into samples in photoemission electron microscopes},}\
  }\href {https://doi.org/10.1063/1.3495967} {\bibfield  {journal} {\bibinfo
  {journal} {Review of Scientific Instruments}\ }\textbf {\bibinfo {volume}
  {81}},\ \bibinfo {pages} {113707} (\bibinfo {year} {2010})}\BibitemShut
  {NoStop}%
\bibitem [{\citenamefont {Le~Guyader}\ \emph {et~al.}(2015)\citenamefont
  {Le~Guyader}, \citenamefont {Savoini}, \citenamefont {El~Moussaoui},
  \citenamefont {Buzzi}, \citenamefont {Tsukamoto}, \citenamefont {Itoh},
  \citenamefont {Kirilyuk}, \citenamefont {Rasing}, \citenamefont {Kimel},\
  and\ \citenamefont {Nolting}}]{LeGuyader2015}%
  \BibitemOpen
  \bibfield  {author} {\bibinfo {author} {\bibfnamefont {L.}~\bibnamefont
  {Le~Guyader}}, \bibinfo {author} {\bibfnamefont {M.}~\bibnamefont {Savoini}},
  \bibinfo {author} {\bibfnamefont {S.}~\bibnamefont {El~Moussaoui}}, \bibinfo
  {author} {\bibfnamefont {M.}~\bibnamefont {Buzzi}}, \bibinfo {author}
  {\bibfnamefont {A.}~\bibnamefont {Tsukamoto}}, \bibinfo {author}
  {\bibfnamefont {A.}~\bibnamefont {Itoh}}, \bibinfo {author} {\bibfnamefont
  {A.}~\bibnamefont {Kirilyuk}}, \bibinfo {author} {\bibfnamefont
  {T.}~\bibnamefont {Rasing}}, \bibinfo {author} {\bibfnamefont {A.~V.}\
  \bibnamefont {Kimel}},\ and\ \bibinfo {author} {\bibfnamefont
  {F.}~\bibnamefont {Nolting}},\ }\bibfield  {title} {\enquote {\bibinfo
  {title} {Nanoscale sub-100 picosecond all-optical magnetization switching in
  {GdFeCo} microstructures},}\ }\href {https://doi.org/10.1038/ncomms6839}
  {\bibfield  {journal} {\bibinfo  {journal} {Nature Communications}\ }\textbf
  {\bibinfo {volume} {6}},\ \bibinfo {pages} {5839} (\bibinfo {year}
  {2015})}\BibitemShut {NoStop}%
\bibitem [{\citenamefont {Gierster}\ \emph {et~al.}(2015)\citenamefont
  {Gierster}, \citenamefont {Ünal}, \citenamefont {Pape}, \citenamefont
  {Radu},\ and\ \citenamefont {Kronast}}]{Gierster2015}%
  \BibitemOpen
  \bibfield  {author} {\bibinfo {author} {\bibfnamefont {L.}~\bibnamefont
  {Gierster}}, \bibinfo {author} {\bibfnamefont {A.~A.}\ \bibnamefont {Ünal}},
  \bibinfo {author} {\bibfnamefont {L.}~\bibnamefont {Pape}}, \bibinfo {author}
  {\bibfnamefont {F.}~\bibnamefont {Radu}},\ and\ \bibinfo {author}
  {\bibfnamefont {F.}~\bibnamefont {Kronast}},\ }\bibfield  {title} {\enquote
  {\bibinfo {title} {Laser induced magnetization switching in a {TbFeCo}
  ferrimagnetic thin film: discerning the impact of dipolar fields, laser
  heating and laser helicity by {XPEEM}},}\ }\href
  {https://doi.org/10.1016/j.ultramic.2015.05.016} {\bibfield  {journal}
  {\bibinfo  {journal} {Ultramicroscopy}\ }\bibinfo {series} {Special {Issue}:
  {LEEM}-{PEEM} 9},\ \textbf {\bibinfo {volume} {159}},\ \bibinfo {pages}
  {508--512} (\bibinfo {year} {2015})}\BibitemShut {NoStop}%
\bibitem [{\citenamefont {Bonetti}\ \emph {et~al.}(2015)\citenamefont
  {Bonetti}, \citenamefont {Kukreja}, \citenamefont {Chen}, \citenamefont
  {Spoddig}, \citenamefont {Ollefs}, \citenamefont {Schöppner}, \citenamefont
  {Meckenstock}, \citenamefont {Ney}, \citenamefont {Pinto}, \citenamefont
  {Houanche}, \citenamefont {Frisch}, \citenamefont {Stöhr}, \citenamefont
  {Dürr},\ and\ \citenamefont {Ohldag}}]{Bonetti2015}%
  \BibitemOpen
  \bibfield  {author} {\bibinfo {author} {\bibfnamefont {S.}~\bibnamefont
  {Bonetti}}, \bibinfo {author} {\bibfnamefont {R.}~\bibnamefont {Kukreja}},
  \bibinfo {author} {\bibfnamefont {Z.}~\bibnamefont {Chen}}, \bibinfo {author}
  {\bibfnamefont {D.}~\bibnamefont {Spoddig}}, \bibinfo {author} {\bibfnamefont
  {K.}~\bibnamefont {Ollefs}}, \bibinfo {author} {\bibfnamefont
  {C.}~\bibnamefont {Schöppner}}, \bibinfo {author} {\bibfnamefont
  {R.}~\bibnamefont {Meckenstock}}, \bibinfo {author} {\bibfnamefont
  {A.}~\bibnamefont {Ney}}, \bibinfo {author} {\bibfnamefont {J.}~\bibnamefont
  {Pinto}}, \bibinfo {author} {\bibfnamefont {R.}~\bibnamefont {Houanche}},
  \bibinfo {author} {\bibfnamefont {J.}~\bibnamefont {Frisch}}, \bibinfo
  {author} {\bibfnamefont {J.}~\bibnamefont {Stöhr}}, \bibinfo {author}
  {\bibfnamefont {H.~A.}\ \bibnamefont {Dürr}},\ and\ \bibinfo {author}
  {\bibfnamefont {H.}~\bibnamefont {Ohldag}},\ }\bibfield  {title} {\enquote
  {\bibinfo {title} {Microwave soft x-ray microscopy for nanoscale
  magnetization dynamics in the 5–10 {GHz} frequency range},}\ }\href
  {https://doi.org/10.1063/1.4930007} {\bibfield  {journal} {\bibinfo
  {journal} {Review of Scientific Instruments}\ }\textbf {\bibinfo {volume}
  {86}},\ \bibinfo {pages} {093703} (\bibinfo {year} {2015})}\BibitemShut
  {NoStop}%
\bibitem [{\citenamefont {Weigand}\ \emph {et~al.}(2022)\citenamefont
  {Weigand}, \citenamefont {Wintz}, \citenamefont {Gräfe}, \citenamefont
  {Noske}, \citenamefont {Stoll}, \citenamefont {Van~Waeyenberge},\ and\
  \citenamefont {Schütz}}]{Weigand2022}%
  \BibitemOpen
  \bibfield  {author} {\bibinfo {author} {\bibfnamefont {M.}~\bibnamefont
  {Weigand}}, \bibinfo {author} {\bibfnamefont {S.}~\bibnamefont {Wintz}},
  \bibinfo {author} {\bibfnamefont {J.}~\bibnamefont {Gräfe}}, \bibinfo
  {author} {\bibfnamefont {M.}~\bibnamefont {Noske}}, \bibinfo {author}
  {\bibfnamefont {H.}~\bibnamefont {Stoll}}, \bibinfo {author} {\bibfnamefont
  {B.}~\bibnamefont {Van~Waeyenberge}},\ and\ \bibinfo {author} {\bibfnamefont
  {G.}~\bibnamefont {Schütz}},\ }\bibfield  {title} {\enquote {\bibinfo
  {title} {{TimeMaxyne}: {A} {Shot}-{Noise} {Limited}, {Time}-{Resolved}
  {Pump}-and-{Probe} {Acquisition} {System} {Capable} of 50 {GHz} {Frequencies}
  for {Synchrotron}-{Based} {X}-ray {Microscopy}},}\ }\href
  {https://doi.org/10.3390/cryst12081029} {\bibfield  {journal} {\bibinfo
  {journal} {Crystals}\ }\textbf {\bibinfo {volume} {12}},\ \bibinfo {pages}
  {1029} (\bibinfo {year} {2022})}\BibitemShut {NoStop}%
\bibitem [{\citenamefont {Stanciu}\ \emph {et~al.}(2007)\citenamefont
  {Stanciu}, \citenamefont {Hansteen}, \citenamefont {Kimel}, \citenamefont
  {Kirilyuk}, \citenamefont {Tsukamoto}, \citenamefont {Itoh},\ and\
  \citenamefont {Rasing}}]{Stanciu2007}%
  \BibitemOpen
  \bibfield  {author} {\bibinfo {author} {\bibfnamefont {C.~D.}\ \bibnamefont
  {Stanciu}}, \bibinfo {author} {\bibfnamefont {F.}~\bibnamefont {Hansteen}},
  \bibinfo {author} {\bibfnamefont {A.~V.}\ \bibnamefont {Kimel}}, \bibinfo
  {author} {\bibfnamefont {A.}~\bibnamefont {Kirilyuk}}, \bibinfo {author}
  {\bibfnamefont {A.}~\bibnamefont {Tsukamoto}}, \bibinfo {author}
  {\bibfnamefont {A.}~\bibnamefont {Itoh}},\ and\ \bibinfo {author}
  {\bibfnamefont {T.}~\bibnamefont {Rasing}},\ }\bibfield  {title} {\enquote
  {\bibinfo {title} {All-{Optical} {Magnetic} {Recording} with {Circularly}
  {Polarized} {Light}},}\ }\href
  {https://doi.org/10.1103/PhysRevLett.99.047601} {\bibfield  {journal}
  {\bibinfo  {journal} {Physical Review Letters}\ }\textbf {\bibinfo {volume}
  {99}},\ \bibinfo {pages} {047601} (\bibinfo {year} {2007})}\BibitemShut
  {NoStop}%
\bibitem [{\citenamefont {Radu}\ \emph {et~al.}(2011)\citenamefont {Radu},
  \citenamefont {Vahaplar}, \citenamefont {Stamm}, \citenamefont {Kachel},
  \citenamefont {Pontius}, \citenamefont {Dürr}, \citenamefont {Ostler},
  \citenamefont {Barker}, \citenamefont {Evans}, \citenamefont {Chantrell},
  \citenamefont {Tsukamoto}, \citenamefont {Itoh}, \citenamefont {Kirilyuk},
  \citenamefont {Rasing},\ and\ \citenamefont {Kimel}}]{Radu2011}%
  \BibitemOpen
  \bibfield  {author} {\bibinfo {author} {\bibfnamefont {I.}~\bibnamefont
  {Radu}}, \bibinfo {author} {\bibfnamefont {K.}~\bibnamefont {Vahaplar}},
  \bibinfo {author} {\bibfnamefont {C.}~\bibnamefont {Stamm}}, \bibinfo
  {author} {\bibfnamefont {T.}~\bibnamefont {Kachel}}, \bibinfo {author}
  {\bibfnamefont {N.}~\bibnamefont {Pontius}}, \bibinfo {author} {\bibfnamefont
  {H.~A.}\ \bibnamefont {Dürr}}, \bibinfo {author} {\bibfnamefont {T.~A.}\
  \bibnamefont {Ostler}}, \bibinfo {author} {\bibfnamefont {J.}~\bibnamefont
  {Barker}}, \bibinfo {author} {\bibfnamefont {R.~F.~L.}\ \bibnamefont
  {Evans}}, \bibinfo {author} {\bibfnamefont {R.~W.}\ \bibnamefont
  {Chantrell}}, \bibinfo {author} {\bibfnamefont {A.}~\bibnamefont
  {Tsukamoto}}, \bibinfo {author} {\bibfnamefont {A.}~\bibnamefont {Itoh}},
  \bibinfo {author} {\bibfnamefont {A.}~\bibnamefont {Kirilyuk}}, \bibinfo
  {author} {\bibfnamefont {T.}~\bibnamefont {Rasing}},\ and\ \bibinfo {author}
  {\bibfnamefont {A.~V.}\ \bibnamefont {Kimel}},\ }\bibfield  {title} {\enquote
  {\bibinfo {title} {Transient ferromagnetic-like state mediating ultrafast
  reversal of antiferromagnetically coupled spins},}\ }\href
  {https://doi.org/10.1038/nature09901} {\bibfield  {journal} {\bibinfo
  {journal} {Nature}\ }\textbf {\bibinfo {volume} {472}},\ \bibinfo {pages}
  {205--208} (\bibinfo {year} {2011})}\BibitemShut {NoStop}%
\bibitem [{\citenamefont {Bergeard}\ \emph {et~al.}(2014)\citenamefont
  {Bergeard}, \citenamefont {L{\'o}pez-Flores}, \citenamefont {Halt{\'e}},
  \citenamefont {Hehn}, \citenamefont {Stamm}, \citenamefont {Pontius},
  \citenamefont {Beaurepaire},\ and\ \citenamefont {Boeglin}}]{Bergeard2014}%
  \BibitemOpen
  \bibfield  {author} {\bibinfo {author} {\bibfnamefont {N.}~\bibnamefont
  {Bergeard}}, \bibinfo {author} {\bibfnamefont {V.}~\bibnamefont
  {L{\'o}pez-Flores}}, \bibinfo {author} {\bibfnamefont {V.}~\bibnamefont
  {Halt{\'e}}}, \bibinfo {author} {\bibfnamefont {M.}~\bibnamefont {Hehn}},
  \bibinfo {author} {\bibfnamefont {C.}~\bibnamefont {Stamm}}, \bibinfo
  {author} {\bibfnamefont {N.}~\bibnamefont {Pontius}}, \bibinfo {author}
  {\bibfnamefont {E.}~\bibnamefont {Beaurepaire}},\ and\ \bibinfo {author}
  {\bibfnamefont {C.}~\bibnamefont {Boeglin}},\ }\bibfield  {title} {\enquote
  {\bibinfo {title} {Ultrafast angular momentum transfer in multisublattice
  ferrimagnets},}\ }\href {https://doi.org/10.1038/ncomms4466} {\bibfield
  {journal} {\bibinfo  {journal} {Nature Communications}\ }\textbf {\bibinfo
  {volume} {5}},\ \bibinfo {pages} {3466} (\bibinfo {year} {2014})}\BibitemShut
  {NoStop}%
\bibitem [{\citenamefont {Radu}\ \emph {et~al.}(2015)\citenamefont {Radu},
  \citenamefont {Stamm}, \citenamefont {Eschenlohr}, \citenamefont {Radu},
  \citenamefont {Abrudan}, \citenamefont {Vahaplar}, \citenamefont {Kachel},
  \citenamefont {Pontius}, \citenamefont {Mitzner}, \citenamefont {Holldack},
  \citenamefont {F{\"o}hlisch}, \citenamefont {Ostler}, \citenamefont
  {Mentink}, \citenamefont {Evans}, \citenamefont {Chantrell}, \citenamefont
  {Tsukamoto}, \citenamefont {Itoh}, \citenamefont {Kirilyuk}, \citenamefont
  {Kimel},\ and\ \citenamefont {Rasing}}]{Radu2015}%
  \BibitemOpen
  \bibfield  {author} {\bibinfo {author} {\bibfnamefont {I.}~\bibnamefont
  {Radu}}, \bibinfo {author} {\bibfnamefont {C.}~\bibnamefont {Stamm}},
  \bibinfo {author} {\bibfnamefont {A.}~\bibnamefont {Eschenlohr}}, \bibinfo
  {author} {\bibfnamefont {F.}~\bibnamefont {Radu}}, \bibinfo {author}
  {\bibfnamefont {R.}~\bibnamefont {Abrudan}}, \bibinfo {author} {\bibfnamefont
  {K.}~\bibnamefont {Vahaplar}}, \bibinfo {author} {\bibfnamefont
  {T.}~\bibnamefont {Kachel}}, \bibinfo {author} {\bibfnamefont
  {N.}~\bibnamefont {Pontius}}, \bibinfo {author} {\bibfnamefont
  {R.}~\bibnamefont {Mitzner}}, \bibinfo {author} {\bibfnamefont
  {K.}~\bibnamefont {Holldack}}, \bibinfo {author} {\bibfnamefont
  {A.}~\bibnamefont {F{\"o}hlisch}}, \bibinfo {author} {\bibfnamefont {T.~A.}\
  \bibnamefont {Ostler}}, \bibinfo {author} {\bibfnamefont {J.~H.}\
  \bibnamefont {Mentink}}, \bibinfo {author} {\bibfnamefont {R.~F.~L.}\
  \bibnamefont {Evans}}, \bibinfo {author} {\bibfnamefont {R.~W.}\ \bibnamefont
  {Chantrell}}, \bibinfo {author} {\bibfnamefont {A.}~\bibnamefont
  {Tsukamoto}}, \bibinfo {author} {\bibfnamefont {A.}~\bibnamefont {Itoh}},
  \bibinfo {author} {\bibfnamefont {A.}~\bibnamefont {Kirilyuk}}, \bibinfo
  {author} {\bibfnamefont {A.~V.}\ \bibnamefont {Kimel}},\ and\ \bibinfo
  {author} {\bibfnamefont {T.}~\bibnamefont {Rasing}},\ }\bibfield  {title}
  {\enquote {\bibinfo {title} {Ultrafast and distinct spin dynamics in magnetic
  alloys},}\ }\href {https://doi.org/10.1142/S2010324715500046} {\bibfield
  {journal} {\bibinfo  {journal} {SPIN}\ }\textbf {\bibinfo {volume} {05}},\
  \bibinfo {pages} {1550004} (\bibinfo {year} {2015})}\BibitemShut {NoStop}%
\bibitem [{\citenamefont {Hennecke}\ \emph {et~al.}(2019)\citenamefont
  {Hennecke}, \citenamefont {Radu}, \citenamefont {Abrudan}, \citenamefont
  {Kachel}, \citenamefont {Holldack}, \citenamefont {Mitzner}, \citenamefont
  {Tsukamoto},\ and\ \citenamefont {Eisebitt}}]{Hennecke2019}%
  \BibitemOpen
  \bibfield  {author} {\bibinfo {author} {\bibfnamefont {M.}~\bibnamefont
  {Hennecke}}, \bibinfo {author} {\bibfnamefont {I.}~\bibnamefont {Radu}},
  \bibinfo {author} {\bibfnamefont {R.}~\bibnamefont {Abrudan}}, \bibinfo
  {author} {\bibfnamefont {T.}~\bibnamefont {Kachel}}, \bibinfo {author}
  {\bibfnamefont {K.}~\bibnamefont {Holldack}}, \bibinfo {author}
  {\bibfnamefont {R.}~\bibnamefont {Mitzner}}, \bibinfo {author} {\bibfnamefont
  {A.}~\bibnamefont {Tsukamoto}},\ and\ \bibinfo {author} {\bibfnamefont
  {S.}~\bibnamefont {Eisebitt}},\ }\bibfield  {title} {\enquote {\bibinfo
  {title} {Angular {Momentum} {Flow} {During} {Ultrafast} {Demagnetization} of
  a {Ferrimagnet}},}\ }\href {https://doi.org/10.1103/PhysRevLett.122.157202}
  {\bibfield  {journal} {\bibinfo  {journal} {Physical Review Letters}\
  }\textbf {\bibinfo {volume} {122}},\ \bibinfo {pages} {157202} (\bibinfo
  {year} {2019})}\BibitemShut {NoStop}%
\bibitem [{\citenamefont {Steinbach}\ \emph {et~al.}(2022)\citenamefont
  {Steinbach}, \citenamefont {Stetzuhn}, \citenamefont {Engel}, \citenamefont
  {Atxitia}, \citenamefont {von Korff~Schmising},\ and\ \citenamefont
  {Eisebitt}}]{Steinbach2022}%
  \BibitemOpen
  \bibfield  {author} {\bibinfo {author} {\bibfnamefont {F.}~\bibnamefont
  {Steinbach}}, \bibinfo {author} {\bibfnamefont {N.}~\bibnamefont {Stetzuhn}},
  \bibinfo {author} {\bibfnamefont {D.}~\bibnamefont {Engel}}, \bibinfo
  {author} {\bibfnamefont {U.}~\bibnamefont {Atxitia}}, \bibinfo {author}
  {\bibfnamefont {C.}~\bibnamefont {von Korff~Schmising}},\ and\ \bibinfo
  {author} {\bibfnamefont {S.}~\bibnamefont {Eisebitt}},\ }\bibfield  {title}
  {\enquote {\bibinfo {title} {Accelerating double pulse all-optical
  write/erase cycles in metallic ferrimagnets},}\ }\href
  {https://doi.org/10.1063/5.0080351} {\bibfield  {journal} {\bibinfo
  {journal} {Applied Physics Letters}\ }\textbf {\bibinfo {volume} {120}},\
  \bibinfo {pages} {112406} (\bibinfo {year} {2022})},\ \Eprint
  {https://arxiv.org/abs/https://doi.org/10.1063/5.0080351}
  {https://doi.org/10.1063/5.0080351} \BibitemShut {NoStop}%
\bibitem [{\citenamefont {Nolle}\ \emph {et~al.}(2012)\citenamefont {Nolle},
  \citenamefont {Weigand}, \citenamefont {Audehm}, \citenamefont {Goering},
  \citenamefont {Wiesemann}, \citenamefont {Wolter}, \citenamefont {Nolle},\
  and\ \citenamefont {Schütz}}]{Nolle2012}%
  \BibitemOpen
  \bibfield  {author} {\bibinfo {author} {\bibfnamefont {D.}~\bibnamefont
  {Nolle}}, \bibinfo {author} {\bibfnamefont {M.}~\bibnamefont {Weigand}},
  \bibinfo {author} {\bibfnamefont {P.}~\bibnamefont {Audehm}}, \bibinfo
  {author} {\bibfnamefont {E.}~\bibnamefont {Goering}}, \bibinfo {author}
  {\bibfnamefont {U.}~\bibnamefont {Wiesemann}}, \bibinfo {author}
  {\bibfnamefont {C.}~\bibnamefont {Wolter}}, \bibinfo {author} {\bibfnamefont
  {E.}~\bibnamefont {Nolle}},\ and\ \bibinfo {author} {\bibfnamefont
  {G.}~\bibnamefont {Schütz}},\ }\bibfield  {title} {\enquote {\bibinfo
  {title} {{Unique} characterization possibilities in the ultra high vacuum
  scanning transmission x-ray microscope ({UHV}-{STXM}) “{MAXYMUS}” using a
  rotatable permanent magnetic field up to 0.22 {T}},}\ }\href
  {https://doi.org/10.1063/1.4707747} {\bibfield  {journal} {\bibinfo
  {journal} {Review of Scientific Instruments}\ }\textbf {\bibinfo {volume}
  {83}},\ \bibinfo {pages} {046112} (\bibinfo {year} {2012})}\BibitemShut
  {NoStop}%
\bibitem [{\citenamefont {Vodungbo}\ \emph {et~al.}(2016)\citenamefont
  {Vodungbo}, \citenamefont {Tudu}, \citenamefont {Perron}, \citenamefont
  {Delaunay}, \citenamefont {Müller}, \citenamefont {Berntsen}, \citenamefont
  {Grübel}, \citenamefont {Malinowski}, \citenamefont {Weier}, \citenamefont
  {Gautier}, \citenamefont {Lambert}, \citenamefont {Zeitoun}, \citenamefont
  {Gutt}, \citenamefont {Jal}, \citenamefont {Reid}, \citenamefont {Granitzka},
  \citenamefont {Jaouen}, \citenamefont {Dakovski}, \citenamefont {Moeller},
  \citenamefont {Minitti}, \citenamefont {Mitra}, \citenamefont {Carron},
  \citenamefont {Pfau}, \citenamefont {von Korff~Schmising}, \citenamefont
  {Schneider}, \citenamefont {Eisebitt},\ and\ \citenamefont
  {Lüning}}]{Vodungbo_2016}%
  \BibitemOpen
  \bibfield  {author} {\bibinfo {author} {\bibfnamefont {B.}~\bibnamefont
  {Vodungbo}}, \bibinfo {author} {\bibfnamefont {B.}~\bibnamefont {Tudu}},
  \bibinfo {author} {\bibfnamefont {J.}~\bibnamefont {Perron}}, \bibinfo
  {author} {\bibfnamefont {R.}~\bibnamefont {Delaunay}}, \bibinfo {author}
  {\bibfnamefont {L.}~\bibnamefont {Müller}}, \bibinfo {author} {\bibfnamefont
  {M.~H.}\ \bibnamefont {Berntsen}}, \bibinfo {author} {\bibfnamefont
  {G.}~\bibnamefont {Grübel}}, \bibinfo {author} {\bibfnamefont
  {G.}~\bibnamefont {Malinowski}}, \bibinfo {author} {\bibfnamefont
  {C.}~\bibnamefont {Weier}}, \bibinfo {author} {\bibfnamefont
  {J.}~\bibnamefont {Gautier}}, \bibinfo {author} {\bibfnamefont
  {G.}~\bibnamefont {Lambert}}, \bibinfo {author} {\bibfnamefont
  {P.}~\bibnamefont {Zeitoun}}, \bibinfo {author} {\bibfnamefont
  {C.}~\bibnamefont {Gutt}}, \bibinfo {author} {\bibfnamefont {E.}~\bibnamefont
  {Jal}}, \bibinfo {author} {\bibfnamefont {A.~H.}\ \bibnamefont {Reid}},
  \bibinfo {author} {\bibfnamefont {P.~W.}\ \bibnamefont {Granitzka}}, \bibinfo
  {author} {\bibfnamefont {N.}~\bibnamefont {Jaouen}}, \bibinfo {author}
  {\bibfnamefont {G.~L.}\ \bibnamefont {Dakovski}}, \bibinfo {author}
  {\bibfnamefont {S.}~\bibnamefont {Moeller}}, \bibinfo {author} {\bibfnamefont
  {M.~P.}\ \bibnamefont {Minitti}}, \bibinfo {author} {\bibfnamefont
  {A.}~\bibnamefont {Mitra}}, \bibinfo {author} {\bibfnamefont
  {S.}~\bibnamefont {Carron}}, \bibinfo {author} {\bibfnamefont
  {B.}~\bibnamefont {Pfau}}, \bibinfo {author} {\bibfnamefont {C.}~\bibnamefont
  {von Korff~Schmising}}, \bibinfo {author} {\bibfnamefont {M.}~\bibnamefont
  {Schneider}}, \bibinfo {author} {\bibfnamefont {S.}~\bibnamefont
  {Eisebitt}},\ and\ \bibinfo {author} {\bibfnamefont {J.}~\bibnamefont
  {Lüning}},\ }\bibfield  {title} {\enquote {\bibinfo {title} {Indirect
  excitation of ultrafast demagnetization},}\ }\href
  {https://doi.org/10.1038/srep18970} {\bibfield  {journal} {\bibinfo
  {journal} {Scientific Reports}\ }\textbf {\bibinfo {volume} {6}},\ \bibinfo
  {pages} {18970} (\bibinfo {year} {2016})}\BibitemShut {NoStop}%
\bibitem [{\citenamefont {Kern}\ \emph {et~al.}(2022)\citenamefont {Kern},
  \citenamefont {Pfau}, \citenamefont {Schneider}, \citenamefont {Gerlinger},
  \citenamefont {Deinhart}, \citenamefont {Wittrock}, \citenamefont
  {Sidiropoulos}, \citenamefont {Engel}, \citenamefont {Will}, \citenamefont
  {Günther}, \citenamefont {Litzius}, \citenamefont {Wintz}, \citenamefont
  {Weigand}, \citenamefont {Büttner},\ and\ \citenamefont
  {Eisebitt}}]{Kern2022-2}%
  \BibitemOpen
  \bibfield  {author} {\bibinfo {author} {\bibfnamefont {L.-M.}\ \bibnamefont
  {Kern}}, \bibinfo {author} {\bibfnamefont {B.}~\bibnamefont {Pfau}}, \bibinfo
  {author} {\bibfnamefont {M.}~\bibnamefont {Schneider}}, \bibinfo {author}
  {\bibfnamefont {K.}~\bibnamefont {Gerlinger}}, \bibinfo {author}
  {\bibfnamefont {V.}~\bibnamefont {Deinhart}}, \bibinfo {author}
  {\bibfnamefont {S.}~\bibnamefont {Wittrock}}, \bibinfo {author}
  {\bibfnamefont {T.}~\bibnamefont {Sidiropoulos}}, \bibinfo {author}
  {\bibfnamefont {D.}~\bibnamefont {Engel}}, \bibinfo {author} {\bibfnamefont
  {I.}~\bibnamefont {Will}}, \bibinfo {author} {\bibfnamefont {C.~M.}\
  \bibnamefont {Günther}}, \bibinfo {author} {\bibfnamefont {K.}~\bibnamefont
  {Litzius}}, \bibinfo {author} {\bibfnamefont {S.}~\bibnamefont {Wintz}},
  \bibinfo {author} {\bibfnamefont {M.}~\bibnamefont {Weigand}}, \bibinfo
  {author} {\bibfnamefont {F.}~\bibnamefont {Büttner}},\ and\ \bibinfo
  {author} {\bibfnamefont {S.}~\bibnamefont {Eisebitt}},\ }\bibfield  {title}
  {\enquote {\bibinfo {title} {Tailoring optical excitation to control magnetic
  skyrmion nucleation},}\ }\href {https://doi.org/10.1103/PhysRevB.106.054435}
  {\bibfield  {journal} {\bibinfo  {journal} {Physical Review B}\ }\textbf
  {\bibinfo {volume} {106}},\ \bibinfo {pages} {054435} (\bibinfo {year}
  {2022})}\BibitemShut {NoStop}%
\bibitem [{\citenamefont {von Korff~Schmising}\ \emph
  {et~al.}(2014)\citenamefont {von Korff~Schmising}, \citenamefont {Pfau},
  \citenamefont {Schneider}, \citenamefont {Günther}, \citenamefont
  {Giovannella}, \citenamefont {Perron}, \citenamefont {Vodungbo},
  \citenamefont {Müller}, \citenamefont {Capotondi}, \citenamefont
  {Pedersoli}, \citenamefont {Mahne}, \citenamefont {Lüning},\ and\
  \citenamefont {Eisebitt}}]{vonKorffSchmising2014}%
  \BibitemOpen
  \bibfield  {author} {\bibinfo {author} {\bibfnamefont {C.}~\bibnamefont {von
  Korff~Schmising}}, \bibinfo {author} {\bibfnamefont {B.}~\bibnamefont
  {Pfau}}, \bibinfo {author} {\bibfnamefont {M.}~\bibnamefont {Schneider}},
  \bibinfo {author} {\bibfnamefont {C.}~\bibnamefont {Günther}}, \bibinfo
  {author} {\bibfnamefont {M.}~\bibnamefont {Giovannella}}, \bibinfo {author}
  {\bibfnamefont {J.}~\bibnamefont {Perron}}, \bibinfo {author} {\bibfnamefont
  {B.}~\bibnamefont {Vodungbo}}, \bibinfo {author} {\bibfnamefont
  {L.}~\bibnamefont {Müller}}, \bibinfo {author} {\bibfnamefont
  {F.}~\bibnamefont {Capotondi}}, \bibinfo {author} {\bibfnamefont
  {E.}~\bibnamefont {Pedersoli}}, \bibinfo {author} {\bibfnamefont
  {N.}~\bibnamefont {Mahne}}, \bibinfo {author} {\bibfnamefont
  {J.}~\bibnamefont {Lüning}},\ and\ \bibinfo {author} {\bibfnamefont
  {S.}~\bibnamefont {Eisebitt}},\ }\bibfield  {title} {\enquote {\bibinfo
  {title} {Imaging {Ultrafast} {Demagnetization} {Dynamics} after a {Spatially}
  {Localized} {Optical} {Excitation}},}\ }\href
  {https://doi.org/10.1103/PhysRevLett.112.217203} {\bibfield  {journal}
  {\bibinfo  {journal} {Physical Review Letters}\ }\textbf {\bibinfo {volume}
  {112}},\ \bibinfo {pages} {217203} (\bibinfo {year} {2014})}\BibitemShut
  {NoStop}%
\bibitem [{\citenamefont {Hennecke}\ \emph {et~al.}(2022)\citenamefont
  {Hennecke}, \citenamefont {Schick}, \citenamefont {Sidiropoulos},
  \citenamefont {Willems}, \citenamefont {Heilmann}, \citenamefont {Bock},
  \citenamefont {Ehrentraut}, \citenamefont {Engel}, \citenamefont {Hessing},
  \citenamefont {Pfau}, \citenamefont {Schmidbauer}, \citenamefont {Furchner},
  \citenamefont {Schnuerer}, \citenamefont {von Korff~Schmising},\ and\
  \citenamefont {Eisebitt}}]{Hennecke2022}%
  \BibitemOpen
  \bibfield  {author} {\bibinfo {author} {\bibfnamefont {M.}~\bibnamefont
  {Hennecke}}, \bibinfo {author} {\bibfnamefont {D.}~\bibnamefont {Schick}},
  \bibinfo {author} {\bibfnamefont {T.}~\bibnamefont {Sidiropoulos}}, \bibinfo
  {author} {\bibfnamefont {F.}~\bibnamefont {Willems}}, \bibinfo {author}
  {\bibfnamefont {A.}~\bibnamefont {Heilmann}}, \bibinfo {author}
  {\bibfnamefont {M.}~\bibnamefont {Bock}}, \bibinfo {author} {\bibfnamefont
  {L.}~\bibnamefont {Ehrentraut}}, \bibinfo {author} {\bibfnamefont
  {D.}~\bibnamefont {Engel}}, \bibinfo {author} {\bibfnamefont
  {P.}~\bibnamefont {Hessing}}, \bibinfo {author} {\bibfnamefont
  {B.}~\bibnamefont {Pfau}}, \bibinfo {author} {\bibfnamefont {M.}~\bibnamefont
  {Schmidbauer}}, \bibinfo {author} {\bibfnamefont {A.}~\bibnamefont
  {Furchner}}, \bibinfo {author} {\bibfnamefont {M.}~\bibnamefont {Schnuerer}},
  \bibinfo {author} {\bibfnamefont {C.}~\bibnamefont {von Korff~Schmising}},\
  and\ \bibinfo {author} {\bibfnamefont {S.}~\bibnamefont {Eisebitt}},\
  }\bibfield  {title} {\enquote {\bibinfo {title} {Ultrafast element- and
  depth-resolved magnetization dynamics probed by transverse magneto-optical
  kerr effect spectroscopy in the soft x-ray range},}\ }\href
  {https://doi.org/10.1103/PhysRevResearch.4.L022062} {\bibfield  {journal}
  {\bibinfo  {journal} {Phys. Rev. Research}\ }\textbf {\bibinfo {volume}
  {4}},\ \bibinfo {pages} {L022062} (\bibinfo {year} {2022})}\BibitemShut
  {NoStop}%
\bibitem [{\citenamefont {Chanda}\ \emph {et~al.}(2021)\citenamefont {Chanda},
  \citenamefont {Shoup}, \citenamefont {Schulz}, \citenamefont {Arena},\ and\
  \citenamefont {Srikanth}}]{Chanda2021}%
  \BibitemOpen
  \bibfield  {author} {\bibinfo {author} {\bibfnamefont {A.}~\bibnamefont
  {Chanda}}, \bibinfo {author} {\bibfnamefont {J.~E.}\ \bibnamefont {Shoup}},
  \bibinfo {author} {\bibfnamefont {N.}~\bibnamefont {Schulz}}, \bibinfo
  {author} {\bibfnamefont {D.~A.}\ \bibnamefont {Arena}},\ and\ \bibinfo
  {author} {\bibfnamefont {H.}~\bibnamefont {Srikanth}},\ }\bibfield  {title}
  {\enquote {\bibinfo {title} {Tunable competing magnetic anisotropies and spin
  reconfigurations in ferrimagnetic {Fe} 100 - x {Gd} x alloy films},}\ }\href
  {https://doi.org/10.1103/PhysRevB.104.094404} {\bibfield  {journal} {\bibinfo
   {journal} {Physical Review B}\ }\textbf {\bibinfo {volume} {104}},\ \bibinfo
  {pages} {094404} (\bibinfo {year} {2021})}\BibitemShut {NoStop}%
\bibitem [{\citenamefont {Talapatra}, \citenamefont {Chelvane},\ and\
  \citenamefont {Mohanty}(2018)}]{Talapatra2018}%
  \BibitemOpen
  \bibfield  {author} {\bibinfo {author} {\bibfnamefont {A.}~\bibnamefont
  {Talapatra}}, \bibinfo {author} {\bibfnamefont {J.~A.}\ \bibnamefont
  {Chelvane}},\ and\ \bibinfo {author} {\bibfnamefont {J.}~\bibnamefont
  {Mohanty}},\ }\bibfield  {title} {\enquote {\bibinfo {title} {Tailoring
  magnetic domains in {Gd}-{Fe} thin films},}\ }\href
  {https://doi.org/10.1063/1.5006413} {\bibfield  {journal} {\bibinfo
  {journal} {AIP Advances}\ }\textbf {\bibinfo {volume} {8}},\ \bibinfo {pages}
  {056327} (\bibinfo {year} {2018})}\BibitemShut {NoStop}%
\bibitem [{\citenamefont {Talapatra}\ \emph {et~al.}(2019)\citenamefont
  {Talapatra}, \citenamefont {Arout~Chelvane}, \citenamefont {Satpati},
  \citenamefont {Kumar},\ and\ \citenamefont {Mohanty}}]{Talapatra2019}%
  \BibitemOpen
  \bibfield  {author} {\bibinfo {author} {\bibfnamefont {A.}~\bibnamefont
  {Talapatra}}, \bibinfo {author} {\bibfnamefont {J.}~\bibnamefont
  {Arout~Chelvane}}, \bibinfo {author} {\bibfnamefont {B.}~\bibnamefont
  {Satpati}}, \bibinfo {author} {\bibfnamefont {S.}~\bibnamefont {Kumar}},\
  and\ \bibinfo {author} {\bibfnamefont {J.}~\bibnamefont {Mohanty}},\
  }\bibfield  {title} {\enquote {\bibinfo {title} {Tunable magnetic domains and
  depth resolved microstructure in {Gd}-{Fe} thin films},}\ }\href
  {https://doi.org/10.1016/j.jallcom.2018.10.024} {\bibfield  {journal}
  {\bibinfo  {journal} {Journal of Alloys and Compounds}\ }\textbf {\bibinfo
  {volume} {774}},\ \bibinfo {pages} {1059--1068} (\bibinfo {year}
  {2019})}\BibitemShut {NoStop}%
\bibitem [{\citenamefont {Hansen}\ \emph {et~al.}(1989)\citenamefont {Hansen},
  \citenamefont {Clausen}, \citenamefont {Much}, \citenamefont {Rosenkranz},\
  and\ \citenamefont {Witter}}]{Hansen1989}%
  \BibitemOpen
  \bibfield  {author} {\bibinfo {author} {\bibfnamefont {P.}~\bibnamefont
  {Hansen}}, \bibinfo {author} {\bibfnamefont {C.}~\bibnamefont {Clausen}},
  \bibinfo {author} {\bibfnamefont {G.}~\bibnamefont {Much}}, \bibinfo {author}
  {\bibfnamefont {M.}~\bibnamefont {Rosenkranz}},\ and\ \bibinfo {author}
  {\bibfnamefont {K.}~\bibnamefont {Witter}},\ }\bibfield  {title} {\enquote
  {\bibinfo {title} {Magnetic and magneto‐optical properties of rare‐earth
  transition‐metal alloys containing {Gd}, {Tb}, {Fe}, {Co}},}\ }\href
  {https://doi.org/10.1063/1.343551} {\bibfield  {journal} {\bibinfo  {journal}
  {Journal of Applied Physics}\ }\textbf {\bibinfo {volume} {66}},\ \bibinfo
  {pages} {756--767} (\bibinfo {year} {1989})}\BibitemShut {NoStop}%
\bibitem [{\citenamefont {Yao}\ \emph {et~al.}(2022)\citenamefont {Yao},
  \citenamefont {Steinbach}, \citenamefont {Borchert}, \citenamefont {Schick},
  \citenamefont {Engel}, \citenamefont {Bencivenga}, \citenamefont
  {Mincigrucci}, \citenamefont {Foglia}, \citenamefont {Pedersoli},
  \citenamefont {De~Angelis}, \citenamefont {Pancaldi}, \citenamefont
  {Wehinger}, \citenamefont {Capotondi}, \citenamefont {Masciovecchio},
  \citenamefont {Eisebitt},\ and\ \citenamefont {von
  Korff~Schmising}}]{Yao2022}%
  \BibitemOpen
  \bibfield  {author} {\bibinfo {author} {\bibfnamefont {K.}~\bibnamefont
  {Yao}}, \bibinfo {author} {\bibfnamefont {F.}~\bibnamefont {Steinbach}},
  \bibinfo {author} {\bibfnamefont {M.}~\bibnamefont {Borchert}}, \bibinfo
  {author} {\bibfnamefont {D.}~\bibnamefont {Schick}}, \bibinfo {author}
  {\bibfnamefont {D.}~\bibnamefont {Engel}}, \bibinfo {author} {\bibfnamefont
  {F.}~\bibnamefont {Bencivenga}}, \bibinfo {author} {\bibfnamefont
  {R.}~\bibnamefont {Mincigrucci}}, \bibinfo {author} {\bibfnamefont
  {L.}~\bibnamefont {Foglia}}, \bibinfo {author} {\bibfnamefont
  {E.}~\bibnamefont {Pedersoli}}, \bibinfo {author} {\bibfnamefont
  {D.}~\bibnamefont {De~Angelis}}, \bibinfo {author} {\bibfnamefont
  {M.}~\bibnamefont {Pancaldi}}, \bibinfo {author} {\bibfnamefont
  {B.}~\bibnamefont {Wehinger}}, \bibinfo {author} {\bibfnamefont
  {F.}~\bibnamefont {Capotondi}}, \bibinfo {author} {\bibfnamefont
  {C.}~\bibnamefont {Masciovecchio}}, \bibinfo {author} {\bibfnamefont
  {S.}~\bibnamefont {Eisebitt}},\ and\ \bibinfo {author} {\bibfnamefont
  {C.}~\bibnamefont {von Korff~Schmising}},\ }\bibfield  {title} {\enquote
  {\bibinfo {title} {All-optical switching on the nanometer scale excited and
  probed with femtosecond extreme ultraviolet pulses},}\ }\href
  {https://doi.org/10.1021/acs.nanolett.2c01060} {\bibfield  {journal}
  {\bibinfo  {journal} {Nano Letters}\ }\textbf {\bibinfo {volume} {22}},\
  \bibinfo {pages} {4452--4458} (\bibinfo {year} {2022})}\BibitemShut {NoStop}%
\bibitem [{\citenamefont {Weder}\ \emph {et~al.}(2020)\citenamefont {Weder},
  \citenamefont {{von Korff Schmising}}, \citenamefont {G{\"u}nther},
  \citenamefont {Schneider}, \citenamefont {Engel}, \citenamefont {Hessing},
  \citenamefont {Str{\"u}ber}, \citenamefont {Weigand}, \citenamefont
  {Vodungbo}, \citenamefont {Jal}, \citenamefont {Liu}, \citenamefont {Merhe},
  \citenamefont {Pedersoli}, \citenamefont {Capotondi}, \citenamefont
  {L{\"u}ning}, \citenamefont {Pfau},\ and\ \citenamefont
  {Eisebitt}}]{weder2020}%
  \BibitemOpen
  \bibfield  {author} {\bibinfo {author} {\bibfnamefont {D.}~\bibnamefont
  {Weder}}, \bibinfo {author} {\bibfnamefont {C.}~\bibnamefont {{von Korff
  Schmising}}}, \bibinfo {author} {\bibfnamefont {C.~M.}\ \bibnamefont
  {G{\"u}nther}}, \bibinfo {author} {\bibfnamefont {M.}~\bibnamefont
  {Schneider}}, \bibinfo {author} {\bibfnamefont {D.}~\bibnamefont {Engel}},
  \bibinfo {author} {\bibfnamefont {P.}~\bibnamefont {Hessing}}, \bibinfo
  {author} {\bibfnamefont {C.}~\bibnamefont {Str{\"u}ber}}, \bibinfo {author}
  {\bibfnamefont {M.}~\bibnamefont {Weigand}}, \bibinfo {author} {\bibfnamefont
  {B.}~\bibnamefont {Vodungbo}}, \bibinfo {author} {\bibfnamefont
  {E.}~\bibnamefont {Jal}}, \bibinfo {author} {\bibfnamefont {X.}~\bibnamefont
  {Liu}}, \bibinfo {author} {\bibfnamefont {A.}~\bibnamefont {Merhe}}, \bibinfo
  {author} {\bibfnamefont {E.}~\bibnamefont {Pedersoli}}, \bibinfo {author}
  {\bibfnamefont {F.}~\bibnamefont {Capotondi}}, \bibinfo {author}
  {\bibfnamefont {J.}~\bibnamefont {L{\"u}ning}}, \bibinfo {author}
  {\bibfnamefont {B.}~\bibnamefont {Pfau}},\ and\ \bibinfo {author}
  {\bibfnamefont {S.}~\bibnamefont {Eisebitt}},\ }\bibfield  {title} {\enquote
  {\bibinfo {title} {Transient magnetic gratings on the nanometer scale},}\
  }\href {https://doi.org/10.1063/4.0000017} {\bibfield  {journal} {\bibinfo
  {journal} {Structural Dynamics}\ }\textbf {\bibinfo {volume} {7}},\ \bibinfo
  {pages} {054501} (\bibinfo {year} {2020})}\BibitemShut {NoStop}%
\bibitem [{\citenamefont {Hassdenteufel}\ \emph {et~al.}(2014)\citenamefont
  {Hassdenteufel}, \citenamefont {Schubert}, \citenamefont {Hebler},
  \citenamefont {Schultheiss}, \citenamefont {Fassbender}, \citenamefont
  {Albrecht},\ and\ \citenamefont {Bratschitsch}}]{Hassdenteufel2014}%
  \BibitemOpen
  \bibfield  {author} {\bibinfo {author} {\bibfnamefont {A.}~\bibnamefont
  {Hassdenteufel}}, \bibinfo {author} {\bibfnamefont {C.}~\bibnamefont
  {Schubert}}, \bibinfo {author} {\bibfnamefont {B.}~\bibnamefont {Hebler}},
  \bibinfo {author} {\bibfnamefont {H.}~\bibnamefont {Schultheiss}}, \bibinfo
  {author} {\bibfnamefont {J.}~\bibnamefont {Fassbender}}, \bibinfo {author}
  {\bibfnamefont {M.}~\bibnamefont {Albrecht}},\ and\ \bibinfo {author}
  {\bibfnamefont {R.}~\bibnamefont {Bratschitsch}},\ }\bibfield  {title}
  {\enquote {\bibinfo {title} {All-optical helicity dependent magnetic
  switching in {Tb}-{Fe} thin films with a {MHz} laser oscillator},}\ }\href
  {https://doi.org/10.1364/OE.22.010017} {\bibfield  {journal} {\bibinfo
  {journal} {Optics Express}\ }\textbf {\bibinfo {volume} {22}},\ \bibinfo
  {pages} {10017} (\bibinfo {year} {2014})}\BibitemShut {NoStop}%
\bibitem [{\citenamefont {Möller}\ \emph {et~al.}(2021)\citenamefont
  {Möller}, \citenamefont {Probst}, \citenamefont {Otto}, \citenamefont
  {Stroh}, \citenamefont {Mahn}, \citenamefont {Steil}, \citenamefont
  {Moshnyaga}, \citenamefont {Jansen}, \citenamefont {Steil},\ and\
  \citenamefont {Mathias}}]{Moller2021}%
  \BibitemOpen
  \bibfield  {author} {\bibinfo {author} {\bibfnamefont {C.}~\bibnamefont
  {Möller}}, \bibinfo {author} {\bibfnamefont {H.}~\bibnamefont {Probst}},
  \bibinfo {author} {\bibfnamefont {J.}~\bibnamefont {Otto}}, \bibinfo {author}
  {\bibfnamefont {K.}~\bibnamefont {Stroh}}, \bibinfo {author} {\bibfnamefont
  {C.}~\bibnamefont {Mahn}}, \bibinfo {author} {\bibfnamefont {S.}~\bibnamefont
  {Steil}}, \bibinfo {author} {\bibfnamefont {V.}~\bibnamefont {Moshnyaga}},
  \bibinfo {author} {\bibfnamefont {G.~S.~M.}\ \bibnamefont {Jansen}}, \bibinfo
  {author} {\bibfnamefont {D.}~\bibnamefont {Steil}},\ and\ \bibinfo {author}
  {\bibfnamefont {S.}~\bibnamefont {Mathias}},\ }\bibfield  {title} {\enquote
  {\bibinfo {title} {Ultrafast element-resolved magneto-optics using a
  fiber-laser-driven extreme ultraviolet light source},}\ }\href
  {https://doi.org/10.1063/5.0050883} {\bibfield  {journal} {\bibinfo
  {journal} {Review of Scientific Instruments}\ }\textbf {\bibinfo {volume}
  {92}},\ \bibinfo {pages} {065107} (\bibinfo {year} {2021})}\BibitemShut
  {NoStop}%
\bibitem [{\citenamefont {Jankowiak}\ \emph {et~al.}(2015)\citenamefont
  {Jankowiak}, \citenamefont {Knobloch}, \citenamefont {Goslawski},\ and\
  \citenamefont {Neumann}}]{Jankowiak2015}%
  \BibitemOpen
  \bibfield  {author} {\bibinfo {author} {\bibfnamefont {A.}~\bibnamefont
  {Jankowiak}}, \bibinfo {author} {\bibfnamefont {J.}~\bibnamefont {Knobloch}},
  \bibinfo {author} {\bibfnamefont {P.}~\bibnamefont {Goslawski}},\ and\
  \bibinfo {author} {\bibfnamefont {N.}~\bibnamefont {Neumann}},\ }\href
  {https://doi.org/10.5442/R0001} {\enquote {\bibinfo {title} {Technical
  {{Design Study BESSSY VSR}}},}\ }\bibinfo {type} {Tech. Rep.}\ (\bibinfo
  {institution} {{Helmholtz-Zentrum Berlin f\"ur Materialien und Energie}},\
  \bibinfo {year} {2015})\BibitemShut {NoStop}%
\bibitem [{\citenamefont {Zhou~Hagstr{\"{o}}m}\ \emph
  {et~al.}(2022)\citenamefont {Zhou~Hagstr{\"{o}}m}, \citenamefont {Schneider},
  \citenamefont {Kerber}, \citenamefont {Yaroslavtsev}, \citenamefont
  {Burgos~Parra}, \citenamefont {Beg}, \citenamefont {Lang}, \citenamefont
  {G{\"{u}}nther}, \citenamefont {Seng}, \citenamefont {Kammerbauer},
  \citenamefont {Popescu}, \citenamefont {Pancaldi}, \citenamefont {Neeraj},
  \citenamefont {Polley}, \citenamefont {Jangid}, \citenamefont {Hrkac},
  \citenamefont {Patel}, \citenamefont {Ovcharenko}, \citenamefont {Turenne},
  \citenamefont {Ksenzov}, \citenamefont {Boeglin}, \citenamefont {Baidakova},
  \citenamefont {von Korff~Schmising}, \citenamefont {Borchert}, \citenamefont
  {Vodungbo}, \citenamefont {Chen}, \citenamefont {Luo}, \citenamefont {Radu},
  \citenamefont {M{\"{u}}ller}, \citenamefont {Mart{\'\i}nez~Fl{\'{o}}rez},
  \citenamefont {Philippi-Kobs}, \citenamefont {Riepp}, \citenamefont
  {Roseker}, \citenamefont {Gr{\"{u}}bel}, \citenamefont {Carley},
  \citenamefont {Schlappa}, \citenamefont {Van~Kuiken}, \citenamefont {Gort},
  \citenamefont {Mercadier}, \citenamefont {Agarwal}, \citenamefont
  {Le~Guyader}, \citenamefont {Mercurio}, \citenamefont {Teichmann},
  \citenamefont {Delitz}, \citenamefont {Reich}, \citenamefont {Broers},
  \citenamefont {Hickin}, \citenamefont {Deiter}, \citenamefont {Moore},
  \citenamefont {Rompotis}, \citenamefont {Wang}, \citenamefont {Kane},
  \citenamefont {Venkatesan}, \citenamefont {Meier}, \citenamefont {Pallas},
  \citenamefont {Jezynski}, \citenamefont {Lederer}, \citenamefont {Boukhelef},
  \citenamefont {Szuba}, \citenamefont {Wrona}, \citenamefont {Hauf},
  \citenamefont {Zhu}, \citenamefont {Bergemann}, \citenamefont {Kamil},
  \citenamefont {Kluyver}, \citenamefont {Rosca}, \citenamefont {Spirzewski},
  \citenamefont {Kuster}, \citenamefont {Turcato}, \citenamefont {Lomidze},
  \citenamefont {Samartsev}, \citenamefont {Engelke}, \citenamefont {Porro},
  \citenamefont {Maffessanti}, \citenamefont {Hansen}, \citenamefont
  {Erdinger}, \citenamefont {Fischer}, \citenamefont {Fiorini}, \citenamefont
  {Castoldi}, \citenamefont {Manghisoni}, \citenamefont {Wunderer},
  \citenamefont {Fullerton}, \citenamefont {Shpyrko}, \citenamefont {Gutt},
  \citenamefont {Sanchez-Hanke}, \citenamefont {D{\"{u}}rr}, \citenamefont
  {Iacocca}, \citenamefont {Nembach}, \citenamefont {Keller}, \citenamefont
  {Shaw}, \citenamefont {Silva}, \citenamefont {Kukreja}, \citenamefont
  {Fangohr}, \citenamefont {Eisebitt}, \citenamefont {Kl{\"{a}}ui},
  \citenamefont {Jaouen}, \citenamefont {Scherz}, \citenamefont {Bonetti},\
  and\ \citenamefont {Jal}}]{hagstrom2022}%
  \BibitemOpen
  \bibfield  {author} {\bibinfo {author} {\bibfnamefont {N.}~\bibnamefont
  {Zhou~Hagstr{\"{o}}m}}, \bibinfo {author} {\bibfnamefont {M.}~\bibnamefont
  {Schneider}}, \bibinfo {author} {\bibfnamefont {N.}~\bibnamefont {Kerber}},
  \bibinfo {author} {\bibfnamefont {A.}~\bibnamefont {Yaroslavtsev}}, \bibinfo
  {author} {\bibfnamefont {E.}~\bibnamefont {Burgos~Parra}}, \bibinfo {author}
  {\bibfnamefont {M.}~\bibnamefont {Beg}}, \bibinfo {author} {\bibfnamefont
  {M.}~\bibnamefont {Lang}}, \bibinfo {author} {\bibfnamefont {C.~M.}\
  \bibnamefont {G{\"{u}}nther}}, \bibinfo {author} {\bibfnamefont
  {B.}~\bibnamefont {Seng}}, \bibinfo {author} {\bibfnamefont {F.}~\bibnamefont
  {Kammerbauer}}, \bibinfo {author} {\bibfnamefont {H.}~\bibnamefont
  {Popescu}}, \bibinfo {author} {\bibfnamefont {M.}~\bibnamefont {Pancaldi}},
  \bibinfo {author} {\bibfnamefont {K.}~\bibnamefont {Neeraj}}, \bibinfo
  {author} {\bibfnamefont {D.}~\bibnamefont {Polley}}, \bibinfo {author}
  {\bibfnamefont {R.}~\bibnamefont {Jangid}}, \bibinfo {author} {\bibfnamefont
  {S.~B.}\ \bibnamefont {Hrkac}}, \bibinfo {author} {\bibfnamefont {S.~K.~K.}\
  \bibnamefont {Patel}}, \bibinfo {author} {\bibfnamefont {S.}~\bibnamefont
  {Ovcharenko}}, \bibinfo {author} {\bibfnamefont {D.}~\bibnamefont {Turenne}},
  \bibinfo {author} {\bibfnamefont {D.}~\bibnamefont {Ksenzov}}, \bibinfo
  {author} {\bibfnamefont {C.}~\bibnamefont {Boeglin}}, \bibinfo {author}
  {\bibfnamefont {M.}~\bibnamefont {Baidakova}}, \bibinfo {author}
  {\bibfnamefont {C.}~\bibnamefont {von Korff~Schmising}}, \bibinfo {author}
  {\bibfnamefont {M.}~\bibnamefont {Borchert}}, \bibinfo {author}
  {\bibfnamefont {B.}~\bibnamefont {Vodungbo}}, \bibinfo {author}
  {\bibfnamefont {K.}~\bibnamefont {Chen}}, \bibinfo {author} {\bibfnamefont
  {C.}~\bibnamefont {Luo}}, \bibinfo {author} {\bibfnamefont {F.}~\bibnamefont
  {Radu}}, \bibinfo {author} {\bibfnamefont {L.}~\bibnamefont {M{\"{u}}ller}},
  \bibinfo {author} {\bibfnamefont {M.}~\bibnamefont
  {Mart{\'\i}nez~Fl{\'{o}}rez}}, \bibinfo {author} {\bibfnamefont
  {A.}~\bibnamefont {Philippi-Kobs}}, \bibinfo {author} {\bibfnamefont
  {M.}~\bibnamefont {Riepp}}, \bibinfo {author} {\bibfnamefont
  {W.}~\bibnamefont {Roseker}}, \bibinfo {author} {\bibfnamefont
  {G.}~\bibnamefont {Gr{\"{u}}bel}}, \bibinfo {author} {\bibfnamefont
  {R.}~\bibnamefont {Carley}}, \bibinfo {author} {\bibfnamefont
  {J.}~\bibnamefont {Schlappa}}, \bibinfo {author} {\bibfnamefont {B.~E.}\
  \bibnamefont {Van~Kuiken}}, \bibinfo {author} {\bibfnamefont
  {R.}~\bibnamefont {Gort}}, \bibinfo {author} {\bibfnamefont {L.}~\bibnamefont
  {Mercadier}}, \bibinfo {author} {\bibfnamefont {N.}~\bibnamefont {Agarwal}},
  \bibinfo {author} {\bibfnamefont {L.}~\bibnamefont {Le~Guyader}}, \bibinfo
  {author} {\bibfnamefont {G.}~\bibnamefont {Mercurio}}, \bibinfo {author}
  {\bibfnamefont {M.}~\bibnamefont {Teichmann}}, \bibinfo {author}
  {\bibfnamefont {J.~T.}\ \bibnamefont {Delitz}}, \bibinfo {author}
  {\bibfnamefont {A.}~\bibnamefont {Reich}}, \bibinfo {author} {\bibfnamefont
  {C.}~\bibnamefont {Broers}}, \bibinfo {author} {\bibfnamefont
  {D.}~\bibnamefont {Hickin}}, \bibinfo {author} {\bibfnamefont
  {C.}~\bibnamefont {Deiter}}, \bibinfo {author} {\bibfnamefont
  {J.}~\bibnamefont {Moore}}, \bibinfo {author} {\bibfnamefont
  {D.}~\bibnamefont {Rompotis}}, \bibinfo {author} {\bibfnamefont
  {J.}~\bibnamefont {Wang}}, \bibinfo {author} {\bibfnamefont {D.}~\bibnamefont
  {Kane}}, \bibinfo {author} {\bibfnamefont {S.}~\bibnamefont {Venkatesan}},
  \bibinfo {author} {\bibfnamefont {J.}~\bibnamefont {Meier}}, \bibinfo
  {author} {\bibfnamefont {F.}~\bibnamefont {Pallas}}, \bibinfo {author}
  {\bibfnamefont {T.}~\bibnamefont {Jezynski}}, \bibinfo {author}
  {\bibfnamefont {M.}~\bibnamefont {Lederer}}, \bibinfo {author} {\bibfnamefont
  {D.}~\bibnamefont {Boukhelef}}, \bibinfo {author} {\bibfnamefont
  {J.}~\bibnamefont {Szuba}}, \bibinfo {author} {\bibfnamefont
  {K.}~\bibnamefont {Wrona}}, \bibinfo {author} {\bibfnamefont
  {S.}~\bibnamefont {Hauf}}, \bibinfo {author} {\bibfnamefont {J.}~\bibnamefont
  {Zhu}}, \bibinfo {author} {\bibfnamefont {M.}~\bibnamefont {Bergemann}},
  \bibinfo {author} {\bibfnamefont {E.}~\bibnamefont {Kamil}}, \bibinfo
  {author} {\bibfnamefont {T.}~\bibnamefont {Kluyver}}, \bibinfo {author}
  {\bibfnamefont {R.}~\bibnamefont {Rosca}}, \bibinfo {author} {\bibfnamefont
  {M.}~\bibnamefont {Spirzewski}}, \bibinfo {author} {\bibfnamefont
  {M.}~\bibnamefont {Kuster}}, \bibinfo {author} {\bibfnamefont
  {M.}~\bibnamefont {Turcato}}, \bibinfo {author} {\bibfnamefont
  {D.}~\bibnamefont {Lomidze}}, \bibinfo {author} {\bibfnamefont
  {A.}~\bibnamefont {Samartsev}}, \bibinfo {author} {\bibfnamefont
  {J.}~\bibnamefont {Engelke}}, \bibinfo {author} {\bibfnamefont
  {M.}~\bibnamefont {Porro}}, \bibinfo {author} {\bibfnamefont
  {S.}~\bibnamefont {Maffessanti}}, \bibinfo {author} {\bibfnamefont
  {K.}~\bibnamefont {Hansen}}, \bibinfo {author} {\bibfnamefont
  {F.}~\bibnamefont {Erdinger}}, \bibinfo {author} {\bibfnamefont
  {P.}~\bibnamefont {Fischer}}, \bibinfo {author} {\bibfnamefont
  {C.}~\bibnamefont {Fiorini}}, \bibinfo {author} {\bibfnamefont
  {A.}~\bibnamefont {Castoldi}}, \bibinfo {author} {\bibfnamefont
  {M.}~\bibnamefont {Manghisoni}}, \bibinfo {author} {\bibfnamefont {C.~B.}\
  \bibnamefont {Wunderer}}, \bibinfo {author} {\bibfnamefont {E.~E.}\
  \bibnamefont {Fullerton}}, \bibinfo {author} {\bibfnamefont {O.~G.}\
  \bibnamefont {Shpyrko}}, \bibinfo {author} {\bibfnamefont {C.}~\bibnamefont
  {Gutt}}, \bibinfo {author} {\bibfnamefont {C.}~\bibnamefont {Sanchez-Hanke}},
  \bibinfo {author} {\bibfnamefont {H.~A.}\ \bibnamefont {D{\"{u}}rr}},
  \bibinfo {author} {\bibfnamefont {E.}~\bibnamefont {Iacocca}}, \bibinfo
  {author} {\bibfnamefont {H.~T.}\ \bibnamefont {Nembach}}, \bibinfo {author}
  {\bibfnamefont {M.~W.}\ \bibnamefont {Keller}}, \bibinfo {author}
  {\bibfnamefont {J.~M.}\ \bibnamefont {Shaw}}, \bibinfo {author}
  {\bibfnamefont {T.~J.}\ \bibnamefont {Silva}}, \bibinfo {author}
  {\bibfnamefont {R.}~\bibnamefont {Kukreja}}, \bibinfo {author} {\bibfnamefont
  {H.}~\bibnamefont {Fangohr}}, \bibinfo {author} {\bibfnamefont
  {S.}~\bibnamefont {Eisebitt}}, \bibinfo {author} {\bibfnamefont
  {M.}~\bibnamefont {Kl{\"{a}}ui}}, \bibinfo {author} {\bibfnamefont
  {N.}~\bibnamefont {Jaouen}}, \bibinfo {author} {\bibfnamefont
  {A.}~\bibnamefont {Scherz}}, \bibinfo {author} {\bibfnamefont
  {S.}~\bibnamefont {Bonetti}},\ and\ \bibinfo {author} {\bibfnamefont
  {E.}~\bibnamefont {Jal}},\ }\bibfield  {title} {\enquote {\bibinfo {title}
  {{Megahertz-rate ultrafast X-ray scattering and holographic imaging at the
  European XFEL}},}\ }\href {https://doi.org/10.1107/S1600577522008414}
  {\bibfield  {journal} {\bibinfo  {journal} {Journal of Synchrotron
  Radiation}\ }\textbf {\bibinfo {volume} {29}},\ \bibinfo {pages} {1454--1464}
  (\bibinfo {year} {2022})}\BibitemShut {NoStop}%
\bibitem [{\citenamefont {Barker}\ \emph {et~al.}(2013)\citenamefont {Barker},
  \citenamefont {Atxitia}, \citenamefont {Ostler}, \citenamefont {Hovorka},
  \citenamefont {{Chubykalo-Fesenko}},\ and\ \citenamefont
  {Chantrell}}]{Barker2013}%
  \BibitemOpen
  \bibfield  {author} {\bibinfo {author} {\bibfnamefont {J.}~\bibnamefont
  {Barker}}, \bibinfo {author} {\bibfnamefont {U.}~\bibnamefont {Atxitia}},
  \bibinfo {author} {\bibfnamefont {T.~A.}\ \bibnamefont {Ostler}}, \bibinfo
  {author} {\bibfnamefont {O.}~\bibnamefont {Hovorka}}, \bibinfo {author}
  {\bibfnamefont {O.}~\bibnamefont {{Chubykalo-Fesenko}}},\ and\ \bibinfo
  {author} {\bibfnamefont {R.~W.}\ \bibnamefont {Chantrell}},\ }\bibfield
  {title} {\enquote {\bibinfo {title} {Two-magnon bound state causes ultrafast
  thermally induced magnetisation switching},}\ }\href
  {https://doi.org/10.1038/srep03262} {\bibfield  {journal} {\bibinfo
  {journal} {Scientific Reports}\ }\textbf {\bibinfo {volume} {3}},\ \bibinfo
  {pages} {3262} (\bibinfo {year} {2013})}\BibitemShut {NoStop}%
\bibitem [{\citenamefont {Johnson}\ and\ \citenamefont
  {Christy}(1974)}]{Johnson1974}%
  \BibitemOpen
  \bibfield  {author} {\bibinfo {author} {\bibfnamefont {P.~B.}\ \bibnamefont
  {Johnson}}\ and\ \bibinfo {author} {\bibfnamefont {R.~W.}\ \bibnamefont
  {Christy}},\ }\bibfield  {title} {\enquote {\bibinfo {title} {Optical
  constants of transition metals: Ti, v, cr, mn, fe, co, ni, and pd},}\ }\href
  {https://doi.org/10.1103/PhysRevB.9.5056} {\bibfield  {journal} {\bibinfo
  {journal} {Phys. Rev. B}\ }\textbf {\bibinfo {volume} {9}},\ \bibinfo {pages}
  {5056--5070} (\bibinfo {year} {1974})}\BibitemShut {NoStop}%
\bibitem [{\citenamefont {Gray}(1972)}]{Gray1972}%
  \BibitemOpen
  \bibfield  {author} {\bibinfo {author} {\bibfnamefont {D.~E.}\ \bibnamefont
  {Gray}},\ }\href@noop {} {\emph {\bibinfo {title} {American Institute of
  Physics (AIP). Handbook}}}\ (\bibinfo  {publisher} {New York: McGraw-Hill},\
  \bibinfo {year} {1972})\BibitemShut {NoStop}%
\bibitem [{\citenamefont {Schick}(2021)}]{Schick2021}%
  \BibitemOpen
  \bibfield  {author} {\bibinfo {author} {\bibfnamefont {D.}~\bibnamefont
  {Schick}},\ }\bibfield  {title} {\enquote {\bibinfo {title} {{udkm1Dsim} –
  a {Python} toolbox for simulating {1D} ultrafast dynamics in condensed
  matter},}\ }\href {https://doi.org/10.1016/j.cpc.2021.108031} {\bibfield
  {journal} {\bibinfo  {journal} {Computer Physics Communications}\ }\textbf
  {\bibinfo {volume} {266}},\ \bibinfo {pages} {108031} (\bibinfo {year}
  {2021})}\BibitemShut {NoStop}%
\bibitem [{\citenamefont {K{\"u}ster}, \citenamefont {Bode},\ and\
  \citenamefont {Fritz}(1968)}]{Kuster1968}%
  \BibitemOpen
  \bibfield  {author} {\bibinfo {author} {\bibfnamefont {W.}~\bibnamefont
  {K{\"u}ster}}, \bibinfo {author} {\bibfnamefont {K.~H.}\ \bibnamefont
  {Bode}},\ and\ \bibinfo {author} {\bibfnamefont {W.}~\bibnamefont {Fritz}},\
  }\bibfield  {title} {\enquote {\bibinfo {title} {Beitrag zur {{Messung}} der
  {{W\"armeleitf\"ahigkeit}} von {{Metallen}} im {{Bereich}} von 0 bis
  500\textdegree{{C}}},}\ }\href {https://doi.org/10.1007/BF00751143}
  {\bibfield  {journal} {\bibinfo  {journal} {W\"arme - und
  Stoff\"ubertragung}\ }\textbf {\bibinfo {volume} {1}},\ \bibinfo {pages}
  {129--139} (\bibinfo {year} {1968})}\BibitemShut {NoStop}%
\bibitem [{\citenamefont {Gerlinger}\ \emph {et~al.}(2023)\citenamefont
  {Gerlinger}, \citenamefont {Pfau}, \citenamefont {Hennecke}, \citenamefont
  {Kern}, \citenamefont {Will}, \citenamefont {Noll}, \citenamefont {Weigand},
  \citenamefont {Gräfe}, \citenamefont {Träger}, \citenamefont {Schneider},
  \citenamefont {Günther}, \citenamefont {Engel}, \citenamefont {Schütz},\
  and\ \citenamefont {Eisebitt}}]{Gerlinger2023}%
  \BibitemOpen
  \bibfield  {author} {\bibinfo {author} {\bibfnamefont {K.}~\bibnamefont
  {Gerlinger}}, \bibinfo {author} {\bibfnamefont {B.}~\bibnamefont {Pfau}},
  \bibinfo {author} {\bibfnamefont {M.}~\bibnamefont {Hennecke}}, \bibinfo
  {author} {\bibfnamefont {L.-M.}\ \bibnamefont {Kern}}, \bibinfo {author}
  {\bibfnamefont {I.}~\bibnamefont {Will}}, \bibinfo {author} {\bibfnamefont
  {T.}~\bibnamefont {Noll}}, \bibinfo {author} {\bibfnamefont {M.}~\bibnamefont
  {Weigand}}, \bibinfo {author} {\bibfnamefont {J.}~\bibnamefont {Gräfe}},
  \bibinfo {author} {\bibfnamefont {N.}~\bibnamefont {Träger}}, \bibinfo
  {author} {\bibfnamefont {M.}~\bibnamefont {Schneider}}, \bibinfo {author}
  {\bibfnamefont {C.~M.}\ \bibnamefont {Günther}}, \bibinfo {author}
  {\bibfnamefont {D.}~\bibnamefont {Engel}}, \bibinfo {author} {\bibfnamefont
  {G.}~\bibnamefont {Schütz}},\ and\ \bibinfo {author} {\bibfnamefont
  {S.}~\bibnamefont {Eisebitt}},\ }\href
  {https://doi.org/10.5281/zenodo.7647267} {\enquote {\bibinfo {title} {{Data
  for Pump--probe x-ray microscopy of photo-induced magnetization dynamics at
  MHz repetition rates}},}\ } (\bibinfo {year} {2023}),\ \bibinfo {note}
  {{Zenodo, Dataset, https://doi.org/10.5281/zenodo.7647267}}\BibitemShut
  {NoStop}%
\end{thebibliography}%

\end{document}